\documentclass[sigconf, nonacm]{acmart}

\usepackage{enumitem}
\usepackage{multirow}
\usepackage{graphicx}
\usepackage{dirtytalk}
\usepackage{subcaption}
\usepackage{graphicx}

\usepackage{draftwatermark}

\SetWatermarkText{UNDER REVIEW}
\SetWatermarkScale{0.5}
\SetWatermarkLightness{0.9}

\AtBeginDocument{%
  \providecommand\BibTeX{{%
    \normalfont B\kern-0.5em{\scshape i\kern-0.25em b}\kern-0.8em\TeX}}}

\acmConference[]{}{}{}
\acmYear{}
\setcopyright{none}
\acmDOI{}
\acmISBN{}

\usepackage{booktabs}
\usepackage{multirow}
\usepackage[utf8]{inputenc}
\usepackage{array}
\usepackage[table]{xcolor}
\usepackage{caption}
\usepackage{xspace}

\newcommand{\methodname}{GAUI\xspace}

\newcommand{\revision}[1]{\textcolor{black}{#1}}

\begin{document}


\title [Mind the Gaze]{Mind the Gaze: Improving the Usability of Dwell Input by Adapting Gaze Targets Based on Viewing Distance}
\thanks{Manuscript submitted to an ACM venue}

\author{Omar Namnakani}
\email{o.namnakani.1@research.gla.ac.uk}
\orcid{0000-0002-3803-5781}
\affiliation{%
  \institution{University of Glasgow}
  \country{United Kingdom}
}

\author{Yasmeen Abdrabou}
\email{yasmeen.essam@unibw.de}
\orcid{0000-0002-8895-4997}
\affiliation{%
  \institution{Human-Centered Technologies for Learning,\\ Technical University of Munich}
  \city{Munich}
  \country{Germany}
}

\author{Cristina Fiani}
\orcid{0000-0002-7119-2383}
 \email{c.fiani.1@research.gla.ac.uk}
\affiliation{%
\institution{University of Glasgow}
  \country{United Kingdom}
} 

\author{John H. Williamson}
\orcid{0000-0001-8085-7853}
 \email{johnh.williamson@glasgow.ac.uk}
\affiliation{%
\institution{University of Glasgow}
  \country{United Kingdom}
} 

\author{Mohamed Khamis}
\email{mohamed.khamis@glasgow.ac.uk}
\orcid{0000-0001-7051-5200}
\affiliation{%
  \institution{University of Glasgow}
  \country{Glasgow, United Kingdom}
}  

\renewcommand{\shortauthors}{Namnakani et al.}

\begin{abstract}
    Dwell input shows promise for handheld mobile contexts, but its performance is impacted by target size and viewing distance. 
While fixed target sizes suffice in static setups, in mobile settings, frequent posture changes alter viewing distances, which in turn distort perceived size and hinder dwell performance. 
We address this through \methodname, a Gaze-based Adaptive User Interface that dynamically resizes targets to maximise performance at the given viewing distance. 
\revision{In a two-phased study (N=$24$), \methodname leveraged the strengths of its distance-responsive design, outperforming the large UI static baseline in task time, and being less error-prone than the small UI static baseline. It was rated the most preferred interface overall.}
Participants reflected on using \methodname in six different postures. 
We discuss how their experience is impacted by posture, and propose guidelines for designing context-aware adaptive UIs for dwell interfaces on handheld mobile devices that maximise performance.

\end{abstract}





\begin{teaserfigure}
  \includegraphics[width=\textwidth]{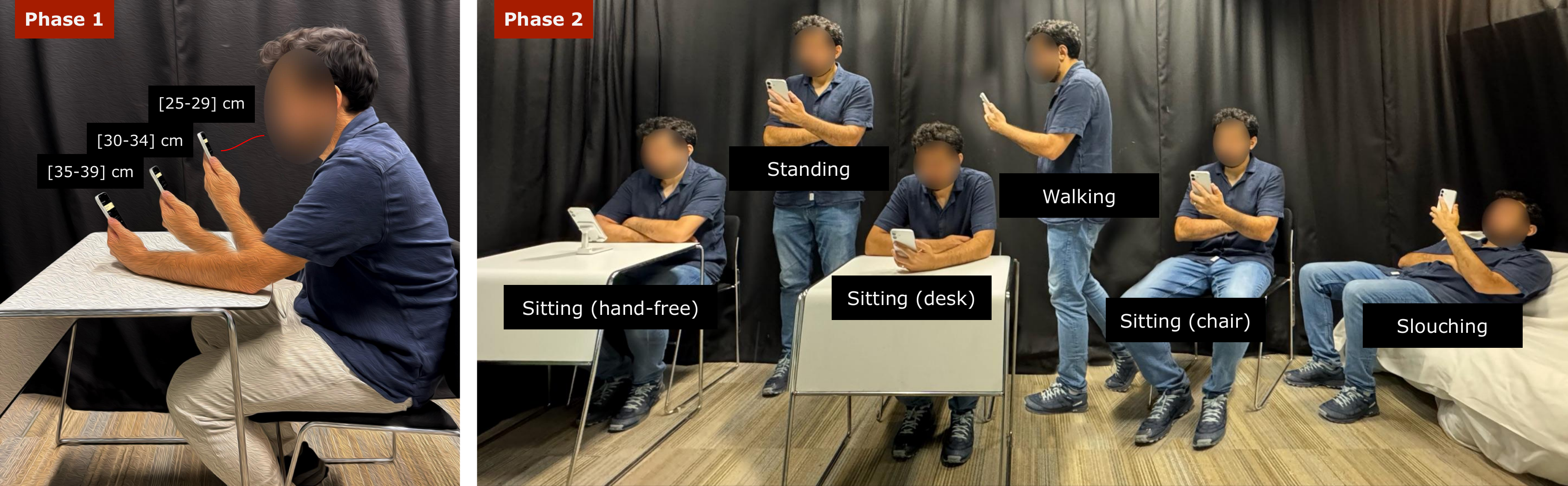}
  \caption{We introduce GAUI, a \revision{\textbf{G}aze-based \textbf{A}daptive \textbf{U}ser \textbf{I}nterface} for dwell input that dynamically resizes UI targets to compensate for the effect of viewing distance on dwell input performance. To examine the proposed UI, we conducted a user study in two phases. In Phase 1, we evaluated the performance of GAUI via a media player application against three static UIs, which served as baselines varying in size: small, medium, and large interfaces, at three different face-to-screen distances (viewing distances). The adaptive UI (\methodname) switches between the three interfaces depending on the distance. The left image shows the setup for Phase 1 of the experiment, where participants sat comfortably on a chair while interacting with the prototype using the four interfaces at various viewing distances. In Phase 2, we adapted the Experience Sampling Method (ESM) to measure the participants' experience and perception of \methodname in a mobile setting while using the adaptive prototype in six different postures: walking, standing, slouching, sitting (hand-free), sitting (chair), and sitting (desk).}
  \label{fig:teaser}
  \Description{Two figures beside each other showing phase 1 on the left side and phase 2 on the right side. In phase 1, it shows one participant interacting with the phone using different face-to-screen distances, namely 25-29 cm, 20-34 cm, and 35-39 cm while sitting at a desk. In phase 2 on the right side, it shows a participant interacting with a phone in different postures, namely: sitting hands-free, standing, sitting at a desk, walking, sitting on a chair, and finally while slouching. }
\end{teaserfigure}

\maketitle

\section{Introduction}
Gaze interaction has been explored on handheld mobile devices~\cite{10.1145/3544548.3580871, 10.1145/3706598.3713092, lei2023dynamicread, 10.1145/3229434.3229452, 10.1145/2168556.2168601}. \revision{It has the potential to be used as an alternative to touch input when its performance degrades due to situational impairments \cite{10.1145/3204493.3208344}, such as when walking encumbered \cite{10.1145/2556288.2557312}, in cold temperatures \cite{goncalves2017tapping}, when wearing gloves \cite{wobbrock2019situationally}, and also when the screen is wet \cite{10.1145/3242969.3243028}.
While voice and hand gestures are other possible input modalities, voice interaction is often infeasible in noisy environments \cite{LaViola2015Multimodal, Chittaro2010}, such as outdoors or on public transport, where mobile devices are frequently used. Voice can be socially unacceptable around strangers \cite{Moorthy2014}, risks privacy leakage \cite{10.1145/3242587.3242603, Moorthy2014}, may disturb people nearby, especially in quiet environments \cite{10.1145/3526114.3558715, Chittaro2010}, and lacks fine-grained control \cite{LaViola2015Multimodal}. While voice could enable eye-free interaction, it could be more attention-demanding and distracting than non-eye-free interfaces \cite{Chittaro2010}.  
Hand gestures can face challenges in dynamic environments \cite{Linardakis2025, Pohekar2025}, are more challenging to perform when hands are occupied, such as when carrying bags, which is common in everyday mobile use \cite{10.1145/3287076, Ng2013, 10.1145/3152771.3156161}, might be embarrassing to perform in public  \cite{khamis2017uist, Chittaro2010}, and may cause arm fatigue (gorilla-arm) \cite{10.1145/2556288.2557130}. Hand- or Motion-based gestures using mobile devices' inertial sensors offer a limited vocabulary \cite{Wang2012, 10.1145/2070481.2070503, 10.1145/1978942.1978971}, may cause hand muscle fatigue, and increase the risk of dropping the phone \cite{10.1145/3544548.3581121}.
On the other hand, prior work suggested that gaze interaction can be fast and convenient \cite{10.1145/1378063.1378122}, intuitive and natural to use \cite{khamis2017uist, 10.1145/2971648.2971679}, and subtle compared to other modalities \cite{10.1145/2800835.2807951}, such as voice input, as others can hardly recognise it in public. It was suggested to become part of everyday interaction with devices such as smartphones \cite{Møllenbach_Hansen_Lillholm_2013, 10.5555/1778331.1778385}. 
Given the current advances in the front-facing cameras and processing power of such devices, recent works explored Dwell time~\cite{10.1145/97243.97246, 10.1145/2168556.2168601}, Pursuits~\cite{10.1145/2493432.2493477, 10.1145/3206505.3206522, 10.1145/2807442.2807499}, and gestures~\cite{10.5555/1778331.1778385, 10.1145/1743666.1743710, Møllenbach_Hansen_Lillholm_2013} for target selections and scrolling when used in the mobile context, while users were sitting and also while on the move~\cite{10.1145/3544548.3580871, lei2023dynamicread}}. \revision{The researchers found that Gestures and Dwell time were more robust than Pursuits while walking~\cite{lei2023dynamicread}, while Dwell time was more favoured for selection than Pursuits and Gestures when walking, and also with an increased number of targets~\cite{10.1145/3544548.3580871}}.
\revision{The user preference for Dwell time in the mobile context could be due to its intuitiveness, ease of learning, the minimal deliberate eye movements required, and also being the gaze counterpart to touch tap~\cite{10.1145/3419249.3420122,10.1145/3544548.3580871, 10.1145/3706598.3713092, 10.1145/2414536.2414609,lei2023dynamicread}}.
\revision{Given the relevance of Dwell time for the mobile context, and that mobile devices are used at various viewing distances (face-to-screen distances) depending on the contexts and postures~\cite{10.1145/3025453.3025794}, recent work by Namnakani et al.~\cite{10.1145/3706598.3713092} explored target sizes for dwell interfaces at viewing distances ranging from 25 to 49 cm, measured as the distance between the user's face and the screen. They found that a target size of 4\textdegree~provides optimal performance across distances.
However, applying their recommended target size in a mobile context is challenging, as viewing distances fluctuate continuously. 
For example, if a designer measured an on-screen target size based on the recommended 4\textdegree~at a viewing distance of $25 cm$, any change in the distance between the user and the device is likely to affect dwell time performance. This occurs because the fixed target size no longer matches the intended perceived size due to changes in viewing distance, making the targets more difficult to select if they look small from farther distances~\cite{10.1145/3706598.3713092, 10.1145/2414536.2414609}}

To address the challenge of degrading performance of dwell time in mobile contexts, we introduce \methodname, a \revision{\textbf{G}aze-based \textbf{A}daptive \textbf{U}ser \textbf{I}nterface} that dynamically rescales targets to preserve their perceptual size as the viewing distance changes. While recent work~\cite{10.1145/3706598.3713092} identified optimal target sizes in a controlled, static setup, our work demonstrates the first adaptive system that applies the ideal target size for maximising the usability of \revision{dwell-based interfaces} in real time on mobile devices. 

\revision{To evaluate our concept, we developed a \methodname-based media player} that dynamically resizes targets depending on the distance. Through a controlled lab experiment, we recruited $24$ participants. \revision{\textbf{In the first phase of the experiment,}} we evaluated the performance of the adaptive UI (\methodname) against three static user interfaces that varied in size (small, medium and large), where participants had to interact with the media player to complete easy and hard tasks. We measured task time, navigation time, two types of task errors (choice of incorrect soundtrack and play/pause control errors), timeouts, workload using NASA-TLX and participants' preference for user interfaces.
\revision{\textbf{In the second phase,}} \revision{participants interacted freely with the \methodname-based media player in six different postures to reflect natural use, where the distance between the user and the device dynamically changes across postures}(see Figure \ref{fig:teaser}).
By adapting the Experience Sampling Method (ESM)~\cite{Larson2014}, we captured their experiences when using \methodname in each posture through the collected responses, logged sensor data, and semi-structured interviews afterwards. 

\revision{Results from \textbf{the first phase} showed that the \textit{adaptive} UI benefited from adapting its target sizes based on viewing distance, drawing on the strength of distance-specific UI, in terms of task and navigation times and errors, while outperforming some static UIs, especially at greater viewing distances.}
\revision{Specifically,} when performing the hard tasks, adaptation significantly reduced task time compared to the large static UI, which required more navigation due to limited screen real estate. It also reduced navigation time, compared to the medium and small static UIs and reduced soundtrack errors compared to the small static UI at farther distances. \revision{Regardless of the viewing distance, the \textit{adaptive} UI significantly reduced the errors from accidental play/pause button activation compared to the small static UI.}
\revision{We found no evidence that the \textit{adaptive} UI increased workload} compared to the static UIs and was also ranked the most preferred interface overall, followed by the medium UI.
\revision{Drawing on participants' experience with \methodname-based media player \textbf{in the second phase}}, they highlighted that their experience varied by posture: while generally positive, it was less favoured during walking. Several suggested adding personalisation features, such as the ability to toggle adaptation on or off depending on context. 

Based on the findings, we derive design guidelines for context-aware, gaze-adaptive UIs. These guidelines show how adaptive resizing can improve target selection for dwell interfaces on handheld mobile devices, while accounting for posture, distance and user preferences.

\section*{Contribution Statement}
We contribute: 1) the first adaptive dwell interface for handheld mobile devices that dynamically preserves perceptual target size across viewing distances; 2) empirical evidence from a two-phase study with 24 participants showing that adaptation improves usability and user preference across diverse postures; and 3) based on the findings, we provide actionable design guidelines for developing context-aware, gaze-adaptive UIs on handheld mobile devices. 

\section{Background and Related Work}
Our work primarily builds on gaze interaction in mobile devices, utilising dwell time, and also on the area of adaptive user interfaces. We conclude the section with a summary of the previous research and our research questions. 

\subsection{Target Sizes for Dwell Interfaces}
Gaze interaction has been studied for a long time, and showed a potential for interaction with smartphones \cite{10.1145/2168556.2168601,10.1145/3229434.3229452}, with Dwell time being the most commonly used technique \cite{10.1145/3544548.3580871, 10.1145/97243.97246}. Dwell time, which is based on eye fixation movements, was introduced to overcome the Midas touch problem by requiring users to fixate on a target for a period of time to perform a selection~\cite{Møllenbach_Hansen_Lillholm_2013}. \revision{Prior work showed that Dwell time holds promise for mobile context, where the technique was favoured over alternative gaze-based techniques that rely on relative eye movements when used for target selection in the mobile context \cite{10.1145/3544548.3580871}, such as Pursuits \cite{10.1145/2493432.2493477, 10.1145/2807442.2807499} and Gestures \cite{10.1145/2168556.2168601, rozado2015controlling, 10.1145/1743666.1743710, 10.5555/1778331.1778385}. Dwell time was also favoured with an increased number of targets \cite{10.1145/3544548.3580871} and was more robust in the mobile context compared to Pursuits when used for scrolling \cite{lei2023dynamicread}. The preference for dwell time is likely due to its intuitiveness, ease of learning, and it also closely resembles clicking or tapping, and requires minimal deliberate eye movement \cite{10.1145/2414536.2414609, 10.1145/123078.128728, 10.1145/3419249.3420122}.}

With regard to target size in dwell input, prior work explored various target sizes for dwell interfaces \cite{10.1145/2414536.2414609, 10.1145/3025453.3025599, niu2021improving, niu2019improving}. However, most of these studies were conducted in non-mobile, non-handheld settings, such as on desktops~\cite{10.1145/2414536.2414609, niu2019improving} and head-mounted displays \cite{esteves2020comparing, 10.1145/3448018.3457998}. \revision{As handheld mobile devices offer only limited screen real estate \cite{10.1145/1378063.1378122, 10.1145/2168556.2168601} compared to desktops and other settings, this limits the applicability of previous findings on target sizes for dwell interfaces \cite{10.1145/3706598.3713092}.} 
Recent work has examined target sizes for dwell interfaces \revision{on handheld mobile devices, taking into consideration the} mobile context \revision{where the distance between the users and mobile devices varies}, investigating the interplay between target size, viewing distance, and target region on tracking accuracy \cite{10.1145/3706598.3713092}. \revision{The researchers explored five target sizes in visual angles (2\textdegree, 3\textdegree, 4\textdegree, and 5\textdegree) and measured their accuracy at five viewing distances between the user and the device, ranging from 25 to 49 cm.} 
The study identified a target size around 4\textdegree~as optimal across viewing distances. \revision{While their recommendation can be applied in static setups where the distance between the user and the screen is fixed,} handheld mobile use is inherently dynamic: as users shift posture or move, \revision{if the target sizes in the user interfaces are fixed in size, and designed for a specific distance, once the distance change,} even an optimal fixed size is perceived differently at different distances, which can degrade accuracy.
In this work, we aim to \revision{optimise target sizes for dwell interfaces in mobile contexts, where viewing distance varies, by exploring the adaptation of user interfaces} that systematically update target sizes in response to changes in viewing distance.

\subsection{Adaptive User Interfaces}
Adaptive user interfaces (AUIs) refer to system-initiated adaptation where UIs dynamically adjust their behaviour in response to contextual factors or user behaviour~\cite{10.1145/3706598.3713367, gulla2015method}. 

Prior work has utilised the adaptive approach to address situational impairments in mobile contexts for touch input, such as when users are walking by scaling targets as their context changes \cite{10.1145/1409240.1409253, 10.1145/2207676.2208662, 10.1145/3706598.3713367}, \revision{to improve mobile touchscreen text entry by adjusting the motor-space locations of keys based on users' behaviour and switching between keyboard models depending on hand posture  \cite{10.1145/2470654.2481386},} or to support the digital well-being of users by discussing the implications of designing an adaptive commitment interface, when visiting Youtube with specific intention by automatically detecting the intention of the user and adapt the user interface accordingly \cite{10.1145/3544548.3580703}. \revision{Junhan et al.~ developed an ability-based design mobile toolkit to enable developers to make their apps responsive to users' situated abilities (touch, gesture, physical activity, and attention), by adapting interface widgets and layouts to better accommodate such abilities~\cite{10.1145/3676524}.}
Other works have utilised eye tracking to explore various forms of UI adaptation, aiming to enhance dwell time selection in explicit gaze-based interactions or to improve privacy when using devices in public spaces for implicit interactions. Examples include adapting the UI to render only words the user is currently reading \cite{10.1145/3338286.3340129}, applying visual masks so that users can see display content only where they are gazing, while the rest of the screen is obscured to prevent bystanders from viewing private data \cite{mti2030045}, or adapting the dwell threshold for gaze-based interfaces based on user performance during interaction \cite{burnham2025effects}. 
Ashmore et al.~looked into using a fisheye lens to magnify the display at the point of the user’s gaze~\cite{ashmore2005efficient}. Other works have investigated expanding the region or the intended targets the user is looking at using a zoom-like tool~\cite{10.1145/1743666.1743702, morimoto2010context, 10.1145/332040.332444, 10.1145/1240624.1240692, 10.1145/3736250, 10.1145/3379155.3391322}.
While such adaptations using eye tracking were explored to protect users' privacy or to enhance dwell selection in interfaces that were not designed for gaze \cite{10.1145/3736250}, our work focuses on using adaptation to enhance dwell selection for gaze-customised targets on handheld mobile devices, taking into account the mobile context. We focus on utilising the adaptive approach to dynamically update target sizes in response to changes in the distance between users and devices while interacting with mobile devices in a mobile context. 
Such contextual factors, when users hold their devices at various distances and in different postures, may negatively impact the usability of dwell interfaces due to tracking inaccuracies when fixating on targets whose perceived size varies with distance. By adapting the user interface, we aim to mitigate the impact of these factors on gaze input, thereby enhancing the usability of the interaction.

\subsection{Summary and Research Questions}
Dwell input shows promising results for the mobile context. However, current UIs use static target sizes. Fixed \revision{target} sizes \revision{may impact dwell time performance} when the distance between the users and the phone varies, as this alters the perceived size of targets and reduces tracking accuracy. 
\revision{Prior work has examined how target size and viewing distance affect dwell performance on mobile devices \cite{10.1145/3706598.3713092}, recommending a target size corresponding to a visual angle of 4\textdegree.} However, handheld use is inherently dynamic: even an `optimal' fixed size is perceived differently across postures and contexts \revision{as the distance varies \cite{10.1145/3025453.3025794, Chittaro2010}, such as when walking, sitting, and slouching}.

Therefore, we investigate adaptive user interfaces that dynamically rescale targets to preserve the intended visual angle across distances. 
We address the following research questions:
\begin{enumerate}[label=\textbf{RQ\arabic*}, leftmargin=1.5cm]
  \item How does dynamically adapting target sizes to viewing distance affect gaze input accuracy and usability in mobile contexts?
  \item How do users perceive and respond to real-time target size changes during interaction, and what design guidelines can be derived for future adaptive gaze interfaces?
\end{enumerate}

\section{Gaze-based Adaptive Mobile User Interface: Concept and Implementation}
For this experiment, we introduce the term \textit{Gaze-based Adaptive User Interfaces} (\methodname) to denote user interfaces designed to compensate for the effect of distance on dwell input performance. 
While other gaze input techniques, such as Pursuits \cite{10.1145/2493432.2493477, 10.1145/2807442.2807499} and Gestures \cite{10.1145/2168556.2168602, 10.1145/3544548.3580871, Møllenbach_Hansen_Lillholm_2013} have been explored in the literature, the Dwell interface remains prevalent as it allows users to point and select, mirroring traditional mouse clicks and touch taps \cite{10.1145/3419249.3420122}. 
Our proposed interface utilises phone sensors to adjust text and target sizes based on the viewing distance between users' faces and their phones. In this section, we explain the concept of Dwell time, \methodname, and the design of our gaze-adaptive media player application.

\subsection{Dwell time} \label{sec:dwell}
To mitigate the Midas touch problem, gaze-enabled interfaces that rely on fixations often incorporate a dwell time to distinguish intentional selections from casual user interface inspection \cite{majaranta2019eye, 10.1145/97243.97246, 10.1145/3419249.3420122}. \revision{In our implementation, we set a dwell threshold of $500\,ms$ for activating control buttons, based on prior work using dwell time in the mobile context \cite{lei2023dynamicread}, for authentication using PIN on desktops \cite{AbdrabouJustGaze2019}, and when evaluating different types of visual feedback for dwell interfaces \cite{10.1145/3419249.3420122}. 
We set a longer dwell time of $1000\,ms$ for activating items in lists, such as soundtrack titles, to give users time to read the texts on the buttons \cite{10.1145/2414536.2414609}.} 
The dwell timer starts when the user's gaze enters the target button, triggering the collection of gaze data with visual feedback in the form of a white stroke outline \revision{highlighting the target under focus \cite{10.1145/968363.968390}}.
The visual feedback remains visible as long as 70\% of the gaze data falls within the target throughout the dwell period and stays until the dwell timer fires \cite{10.1145/3706598.3713092}.
Once the dwell period elapses, the system re-evaluates whether the collected gaze points still fall within the target button. If the 70\% threshold is met, the button's background colour changes to yellow, \revision{providing visual feedback that signals the button’s activation \cite{10.1145/968363.968390}}. 
The threshold approach enhances robustness against noise while providing visual feedback to users, ensuring that minor deviations do not reset the dwell timer as long as 70\% of the gaze points fall within the target at any time during the dwell period. \revision{The visual feedback prior selection when the target is under focus, helps users confirm or cancel their selection as suggested by prior work, and was reported to give participants' confidence over their selection \cite{10.1145/3419249.3420122, Majaranta2006, 10.1145/968363.968390}}.

\subsection{\methodname}
In this section, we explain the concepts associated with \methodname, and present a sample application that applies these concepts to implement a gaze-adaptive media player.

\subsubsection{Target Sizes}
To minimise the negative impact of distance changes on the dwell performance, \methodname enables interface scaling by dynamically resizing UI targets to maintain a consistent perceived visual angle at varying viewing distances throughout the interaction. As preserving the visual angle across viewing distances causes the computed size of the targets to vary, targets may be pushed off-screen and onto the next page. This necessitates consideration of methods to navigate among targets when they overflow to subsequent pages.

\subsubsection{Hysteresis to Prevent Flickering}
To prevent flicker near the viewing distance boundaries when adapting and scaling the UI, \methodname utilises hysteresis by defining a small buffer zone of ±2~cm around each threshold. The UI adapts and target size changes only when the measured viewing distance is beyond this buffer, and reverse changes occur only after crossing the opposite side of the buffer, ensuring smooth, non-jittery transitions.

\subsubsection{Sample Application of \methodname: Media Player} \label{sec:prototype}
We implemented a media player application that uses Dwell time for selection to explore the feasibility and effectiveness of \methodname in mobile settings where distances between users and phones vary. 
The player mimics the user interface of common media players with a list of soundtracks and control buttons, i.e., navigation and play/pause buttons (see Figure \ref{fig:screenshot}). Activating a soundtrack by gaze requires a dwell threshold of 1000\,ms \revision{\cite{10.1145/2414536.2414609}} for the soundtrack to play, whereas the control buttons require a dwell threshold of 500\,ms \revision{\cite{10.1145/3419249.3420122, AbdrabouJustGaze2019, lei2023dynamicread}} (\revision{see Section \ref{sec:dwell} for more details about Dwell time}).

\subsubsection*{Target Sizes}
\revision{We designed the width of items in the soundtrack list to occupy the full-screen width, while we used 4\textdegree~ for heights, which was suggested in prior research \cite{10.1145/3706598.3713092}, as an optimal sweet spot for gaze-enabled interfaces on mobile devices (see Figure \ref{fig:screenshot}).} 
The same work also informed the width and height of control buttons. To fit the three control buttons next to each other in all interfaces for consistency while maintaining tracking accuracy in the horizontal direction, we used a width of 3\textdegree~\cite{10.1145/3706598.3713092}. 
We increased the heights of the control buttons to 5°, as they were placed in the bottom region of the screen, as recommended by the guidelines of prior work \cite{10.1145/3706598.3713092}.

\subsubsection*{Navigation and Control Buttons}
The control buttons consisted of two buttons with right/left arrows for paginated playlist navigation and one for play/pause. 
To control the impact of scrolling speed, we replaced standard scrolling through the playlist with a page-style navigation, where users move left or right to switch between pages of the playlist.
The larger the interface items, the more space they take and thus more pages are needed to navigate all soundtracks in the list. 
\revision{Although the playback controls (next/previous) typically accompany play/pause buttons, we replaced these with page navigation buttons (left/right) for experimental purposes. 
This approach ensured that the control buttons were used more frequently, adding validity to the findings through a higher number of comparable selection events, as one of the tasks required navigating pages to search for soundtracks. This allowed us to better examine the impact of adaptation on the usability of these buttons while keeping the task structure identical across interfaces.}

\section{Evaluation}
We conducted a user study in a quiet room on the university campus in two phases. Phase 1 was for comparing and evaluating the performance of the adaptive UI (\methodname) against static UIs as baselines. In Phase 2, our goal was to capture participants' perceptions of \methodname when used in a mobile context where participants could hold the device in different postures. \revision{While the first phase compared \methodname with static UIs, in which the various viewing distances were controlled to compare the interfaces systematically, this phase aimed to give participants the chance to interact freely with \methodname so we could capture their experience, given the variation in face-to-screen distances and the postures \cite{10.1145/3025453.3025794}}. Our study was approved by the university ethics board (Approval number: 300230214).

\subsection{Tasks} \label{sec:tasks}
In Phase 1 of the experiment, using the media player application, participants were required to play a given soundtrack title shown on the phone's screen. Depending on the task difficulty, as explained in the following section (Section \ref{sec:study_design}), participants were also required to navigate through the playlist to play the given soundtrack.
In Phase 2, participants interacted with the gaze-adaptive media player in six different postures, each for a duration of two minutes. They were instructed and given the freedom to play soundtracks of their choice while navigating the list at their leisure.

\begin{figure}[t]
    \centering
    \includegraphics[width=0.8\linewidth]{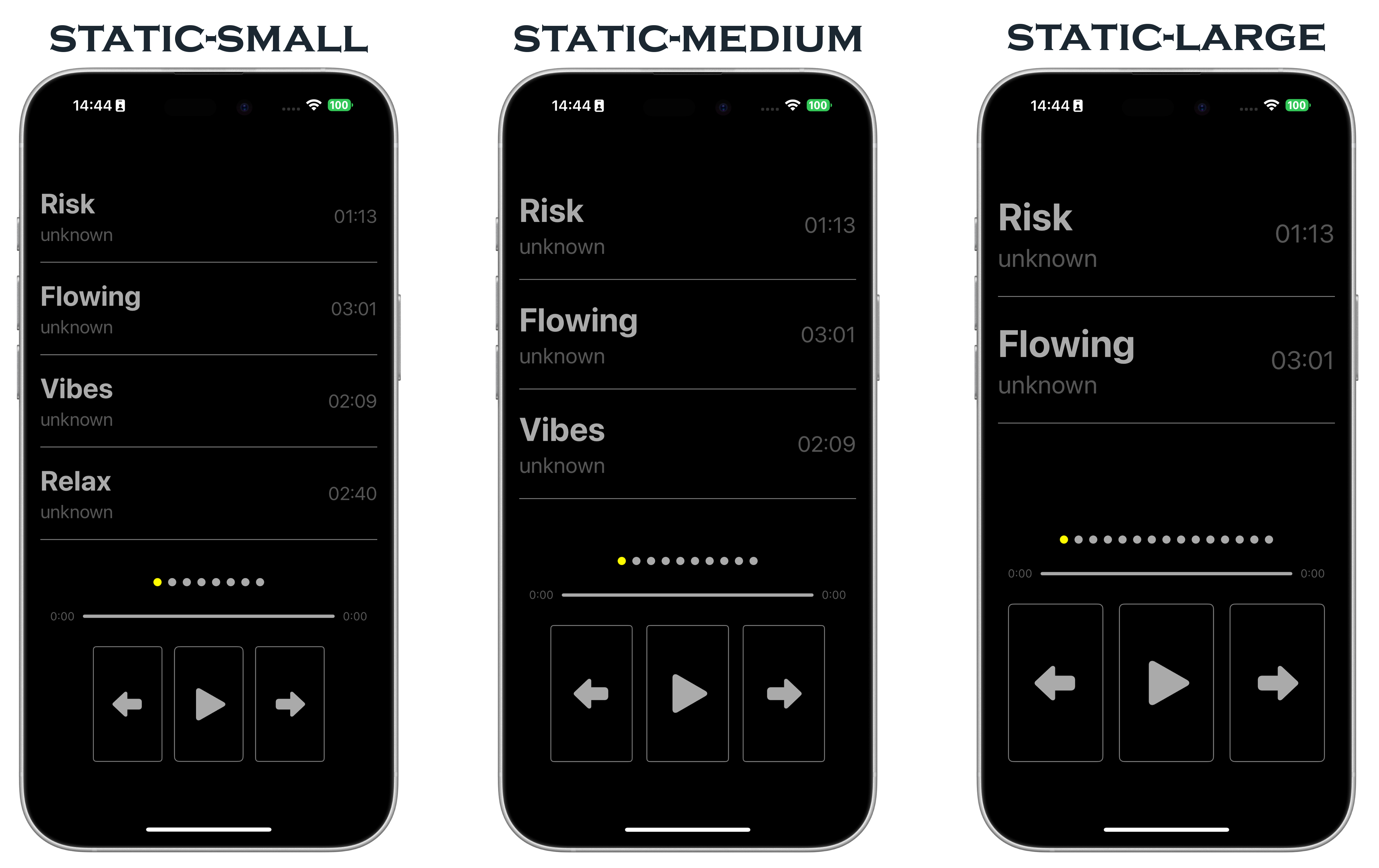}
    \caption{In our experiment, we used four different interface sizes. Besides the three static interfaces shown above, we also created an adaptive interface. The adaptive interface adjusts according to the distance between the participant and the phone to maintain a visual angle of approximately 4° across all distances. At the closest distance, the interface switches to the \textit{static-small}. As the distance increases, the interface size scales up. We calculated the number of soundtracks that could be displayed on each screen based on the size of the interface elements.} 
    \Description{Three smartphone screenshots side by side. The first shows the static-small interface with four soundtracks. The second shows the static-medium interface with three soundtracks. The third shows the static-large interface with two soundtracks. Each interface has playback controls at the bottom. As the interface size increases, all elements scale up and fewer soundtracks fit on the screen.}
    \label{fig:screenshot}
\end{figure}

\subsection{Study Design} \label{sec:study_design}
For phase 1, we designed a within-subject study, which had the following independent variables and their corresponding levels used to evaluate the performance of the Gaze-Adaptive user interfaces (\methodname):
\begin{itemize}
        
    \item \textbf{Face-to-Screen Distance} \{ [25-29] cm, [30-34] cm, [35-39] cm \}: Given the variable nature of mobile use, we defined ranges of distances. Our choice of ranges was informed by previous work of Namnakani et al.~\cite{10.1145/3706598.3713092} and Huang et al.~\cite{10.1145/3025453.3025794}, where the former defined distance ranges when exploring target sizes, and the latter reported that people typically hold smartphones 25–40 cm from their face depending on postures.
    
    \item \textbf{Interface Type} \{ static-small, static-medium, static-large, adaptive \}: The first three conditions are static baselines that do not adapt to the distance changes, whereas the last condition \emph{adapts} to the distance (see Figure \ref{fig:screenshot}). 
    The sizes of targets for each condition were determined as follows. 
    Based on the visual angles discussed in Section \ref{sec:prototype}, we maintained a width that expands the whole screen and a height of 4° for soundtracks and a dimension of 3°$\times$5° for control buttons in all conditions. 
    The sizes of the \textit{static-small}, \textit{static-medium}, and \textit{static-large} interfaces were consistently equivalent to the corresponding target sizes when measured at median distances of [25–29] cm, [30–34] cm, and [35–39] cm, respectively. 
    The \textit{adaptive} interface (\methodname) was designed to preserve the visual angle as the distance changes by switching between three interface sizes: \textit{static-small, static-medium, and static-large}, depending on the face-to-screen distance at which participants hold their phone.

    \item \textbf{Task Difficulty} \{ easy, hard \}: The easy task required participants to locate and play the required soundtrack, which was randomly placed in the currently visible list on the screen, with no navigation required. 
    In the hard task, participants had to navigate across multiple pages to locate and play the soundtrack, which was randomly chosen from the same page as the 9th soundtrack in the list. 
    This means that in the hard task, the soundtrack to be played was located on the third, fourth, and fifth pages for \textit{static-small}, \textit{static-medium}, and \textit{static-large}, respectively. This was done to ensure that task difficulty was comparable across conditions, as the target’s position in the overall list was constant, while still reflecting the realistic trade-off between target size and the number of navigation required.
\end{itemize}

\subsection{Setup and Apparatus} \label{sec:setup}
In Phase 1 of the experiment (Figure \ref{fig:teaser}), participants sat comfortably on a chair with a desk in front of them. To simulate a natural interaction, participants comfortably held the phone, with the option to rest their arm on the desk as long as they maintained the required Face-to-Screen distances. 
In phase 2, \revision{we did not control nor distribute the viewing distances. Instead,} participants were instructed to hold the phone freely without restriction in the six different postures. \revision{We employed postures as a proxy for natural variation in viewing distances \cite{10.1145/3025453.3025794}, since posture changes naturally produce such variation, which in turn produces changes in perceived sizes, rather than repeatedly instructing participants to adjust their distances during the experiment} (see Figure \ref{fig:teaser}).

The media player was developed using Swift as a native iOS application. 
We ran the experiment using an iPhone 16 Plus. The device features a front-facing camera with a 12MP sensor and an f/1.9 aperture. 
The application consisted of 30 soundtracks downloaded from Pixabay.com, which were randomly ordered in each task. We used single-word titles to minimise the risk of participants forgetting the soundtrack.
Eye tracking was enabled using Eyedid SDK \cite{eyedid}. It is a software-based library that uses the phone's front-facing camera to enable tracking. At 30 frames per second and using the RGB images captured, the library provides real-time gaze estimation data recorded as an (x,y) in the screen coordinates. 
For calculating the distance between the camera and the participant's face, we had two options: the ARKit API in iOS, which was highly accurate, and the distance feature in the Eyedid SDK. 
Since using both at the same time was not possible and we needed the Eyedid SDK for eye tracking, we decided to correct any offsets in Eyedid's distance measure using the measurement by the ARKit API at the beginning of the experiment.

 \begin{figure*}[t]
    \centering
    \includegraphics[width=\linewidth]{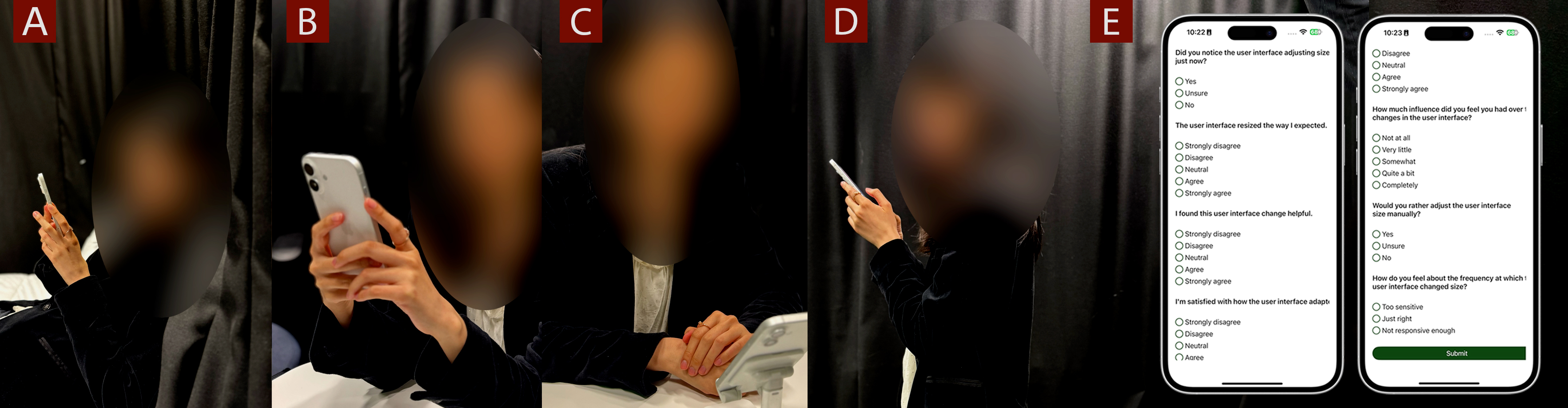}
    \caption{In Phase 2, participants interacted with \methodname in six different postures and responded to in-app questions (\textbf{E}), triggered once an adaptation occurred to reflect on their experience with the adaptive UI in the mobile context. The figure shows a participant exploring the media player while (\textbf{A}) slouching, (\textbf{B}) sitting (desk), (\textbf{C}) sitting (hands-free), and (\textbf{D}) walking.} 
    \Description{Five sub-figures beside each other from A to E from the left side. From A to D, it shows a participant interacting with a phone in different postures. While E shows in-app 8 radio button questions.}
    \label{fig:setup_esm}
\end{figure*}

\subsection{Measures} \label{sec:measures}
To evaluate and compare the performance of the different interfaces in the first phase of the study, our dependent measures were task time, navigation time, task errors, and the number of task timeouts. We define task time as the time it takes to play the correct soundtrack, from when the soundtracks are displayed until the correct soundtrack is played. Navigation time is the average duration participants take to activate a correct navigation button (i.e., left or right buttons). We defined two types of task errors: the first is the number of times participants played incorrect soundtracks before playing the correct one (soundtrack errors), and the second is related to the number of times participants inadvertently activate the play/pause button when aiming to navigate the soundtrack list in the hard task (play/pause control errors). Timeout refers to the number of times participants failed to activate the correct soundtrack within the given time limit \revision{, which is 60 seconds. 
We based our decision on prior work that implemented timeouts for gaze interaction on mobile devices \cite{10.1145/3544548.3580871}. 
Our experiments' tasks require between 1 and 5 selections; applying the 20-second timeout per selection, as in prior work, results in overall task timeouts of 20 to 100 seconds. We chose the middle of that as the overall task timeout limit, i.e., 60 seconds. Note that the timeout is implemented for practical reasons to prevent the experiment from becoming too long. In practice, real-world applications may require shorter timeout thresholds.} 
For each participant, measures were averaged over three trials in each of the 18 experimental conditions. We also collected subjective feedback via NASA-TLX \cite{doi:10.1177/154193120605000909} and Likert-scale questions to measure participants' perceived workload and the usability of the different interfaces.

In the second phase of the experiment, we used an adaptation of the Experience Sampling Method (ESM) \cite{christensen2003practical, Larson2014, 1203750, 10.1145/3544548.3580703} to measure participants' experience and perception of \methodname in six different postures (see Figure \ref{fig:teaser}): while walking, standing, slouching, sitting at a desk + mounted phone (sitting (hands-free)), sitting at a desk + handheld phone (sitting (desk)), and also while sitting on a chair without a desk (sitting (chair)). \revision{This phase allowed participants to interact freely with \methodname-based media player without restriction on the distance they can hold the phone at. We used postures as a proxy for variation in viewing distance \cite{10.1145/3025453.3025794}, enabling us to capture participants' perceptions across varying distances and postures.} 
Using an event-contingent protocol, participants provided real-time feedback immediately after each adaptation event \cite{christensen2003practical, 1203750}. 
This was collected through in-app Likert scale, categorical, and perception-based questions (see Section \ref{sec:phase2} and Figure \ref{fig:setup_esm} (E) for more details about the questions). We concluded with a semi-structured interview \revision{to reflect} on the overall experience \revision{of interacting with \methodname-based media player for target selection when the target sizes adapt based on the viewing distance.}

\subsection{Participants}
\begin{figure}[!t]
    \centering
    \includegraphics[width=0.9\linewidth]{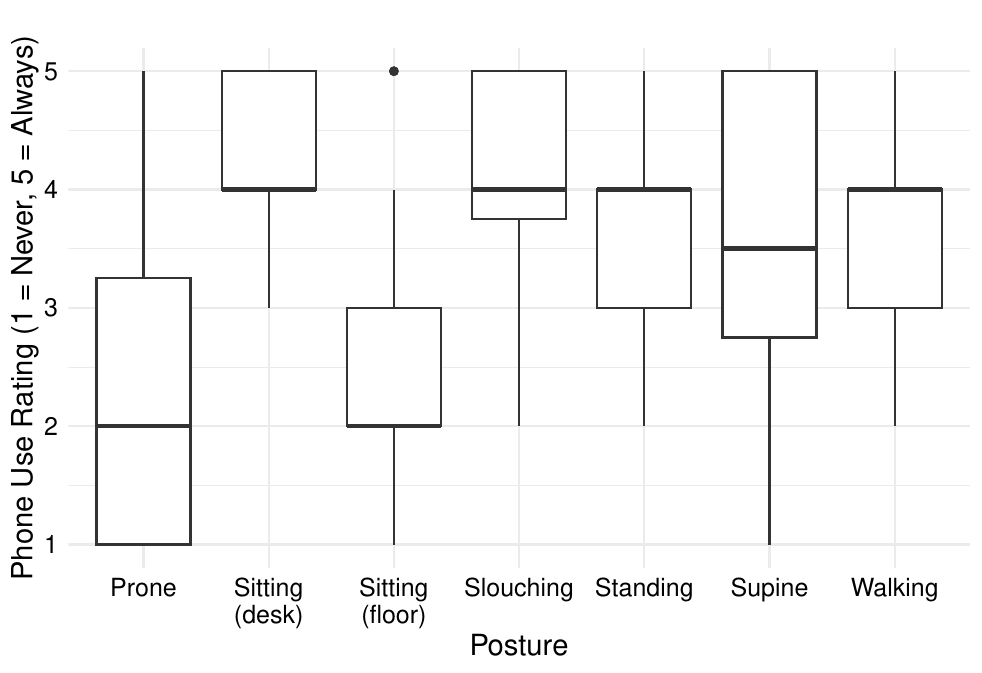}
    \caption{Before the experiment, participants responded to questions using a scale (1 = Never, 2 = Rarely, 3 = Sometimes, 4 = Often, 5 = Always) regarding how frequently they used their phones in various postures. This indicates that participants were familiar with using their phones across various postures, making them suitable for evaluating our interface in realistic contexts.} 
    \Description{Box plot showing the seven postures on the x-axis and the phone usage ranging from 1 = Never and 5 = Always to 5 on the y-axis.}
    \label{fig:phone_use}
\end{figure}

We recruited $24$ participants ($15$ Female, $8$ Males, $1$ Prefer not to say; Mean age = $26.08$, $SD = 6.04$, $min= 18$, $max = 40$), through social media advertising and other means. Participants were compensated with a £10 Amazon voucher. While three participants wore glasses during the experiment, out of the 24 participants, \revision{three reported astigmatism}, four were farsighted, six were nearsighted, \revision{and one reported both farsightedness and astigmatism.} \revision{We asked the participants about the frequency with which they use their phones in various postures (1= Never; 5= Always). They reported more often phone use while walking, standing, slouching, and sitting (see Figure \ref{fig:phone_use})}. They also reported little experience ($M= 1.33$, $SD= 1.61$) with eye tracking (0= no experience; 5= very experienced).

\subsection{Procedure} \label{sec:procedure}
Each participant signed a consent form and answered a survey about their demographics, eye-tracking experience, and Likert scale questions about how often they used their phones in various postures. Then, the experimenter briefed the participants about the study.
Each participant completed two phases of the study: Phase 1 and Phase 2.
The participant completed a 5-point calibration provided by the eye-tracking library once before Phase 1. During calibration, the distance between the camera and the participant's face was set to the middle value of the three Face-to-Screen ranges.

\textbf{In Phase 1}, 
Each participant completed four blocks, each consisting of 18 trials. 
Each block corresponds to a specific interface: \textit{static-small}, \textit{static-medium}, \textit{static-large}, or \textit{adaptive}. 
To familiarise themselves with the task, the participants were allowed to complete two training tasks before proceeding with the experimental blocks, one for each task type (i.e., easy and hard).
\textbf{In each block,} the participant completed three trials for each condition (i.e., each combination of levels from \textit{Task Difficulty} and \textit{Face-to-Screen Distance}).
Blocks and conditions were counterbalanced using a Latin square. 
At the start of each trial, the device displayed an instruction screen informing the participant of the required viewing distance from the phone. This allowed them to adjust the device or their head until the correct distance was achieved. The instruction screen also showed the soundtrack title that the participant needed to play. 
If the participant moved away from the required distance during the trial, an alert prompted them to readjust their position. This alert remained visible until the participant returned to the correct distance.
Once the correct soundtrack was played, a pop-up message confirmed it, and the participant was advanced to the subsequent trial. Each trial ended either when the participant played the correct soundtrack or after a 60-second timeout, indicating a failed trial. 
\textbf{After each block,} we collected feedback using NASA-TLX and Likert-scale questions. 
\textbf{After completing all blocks in Phase 1,} participants ranked their overall preference for \textit{Interface Types} and then proceeded to the second phase.

\textbf{In Phase 2,} participants were allowed to use the \methodname-based media player naturally without restriction in the six postures, as listed in Section \ref{sec:measures}. 
The posture order was counterbalanced. \textbf{In each trial,} participants were shown an instruction screen indicating the required posture.
The trial began once the participant assumed the correct posture and clicked the displayed start button. 
During the task, participants' feedback was collected through ESM in-app questions that appeared as pop-up prompts after the first adaptation occurred in each posture (see Section \ref{sec:phase2} and Figure \ref{fig:setup_esm}). These questions were triggered 30 seconds after the UI adaptation (the event). We employed this duration to allow participants to continue interacting naturally without immediate interruption, enabling them to form a more holistic and reflective understanding of the adaptation.
\textbf{After completing all trials in Phase 2}, we conducted a semi-structured interview as an exit interview to reflect on the participants' experience with \methodname in various postures. 
 
\subsection{Limitations}
In our experiment, we decided to conduct phases 1 and 2 in an indoor environment with no obstacles. While this may not accurately reflect real-world scenarios, where contextual factors such as obstacles or varying lighting conditions exist, we controlled these variables to evaluate the performance of \methodname in isolation.

While ESM is used by researchers to study behaviour over time, we adapted it in Phase 2 to capture participants' experiences with \methodname in a lab setting across different postures. As a result, our findings may not fully reflect everyday use of \methodname, but the method allowed us to gather in-the-moment experiences within the constraints of the lab.

Finally, our findings are based on eye-tracking data collected using the Eyedid library, which also provides the calibration phase. While results may vary with other tracking equipment, in our experiment, all participants and conditions were measured with the same device and tracking library, meaning that any systematic bias introduced should have been applied equally. This ensured that contrasts between conditions were preserved, even if absolute values of fixation might differ when using different hardware. 

\section{Results}

\begin{figure}[!t]
  \centering
    \includegraphics[width=\linewidth]{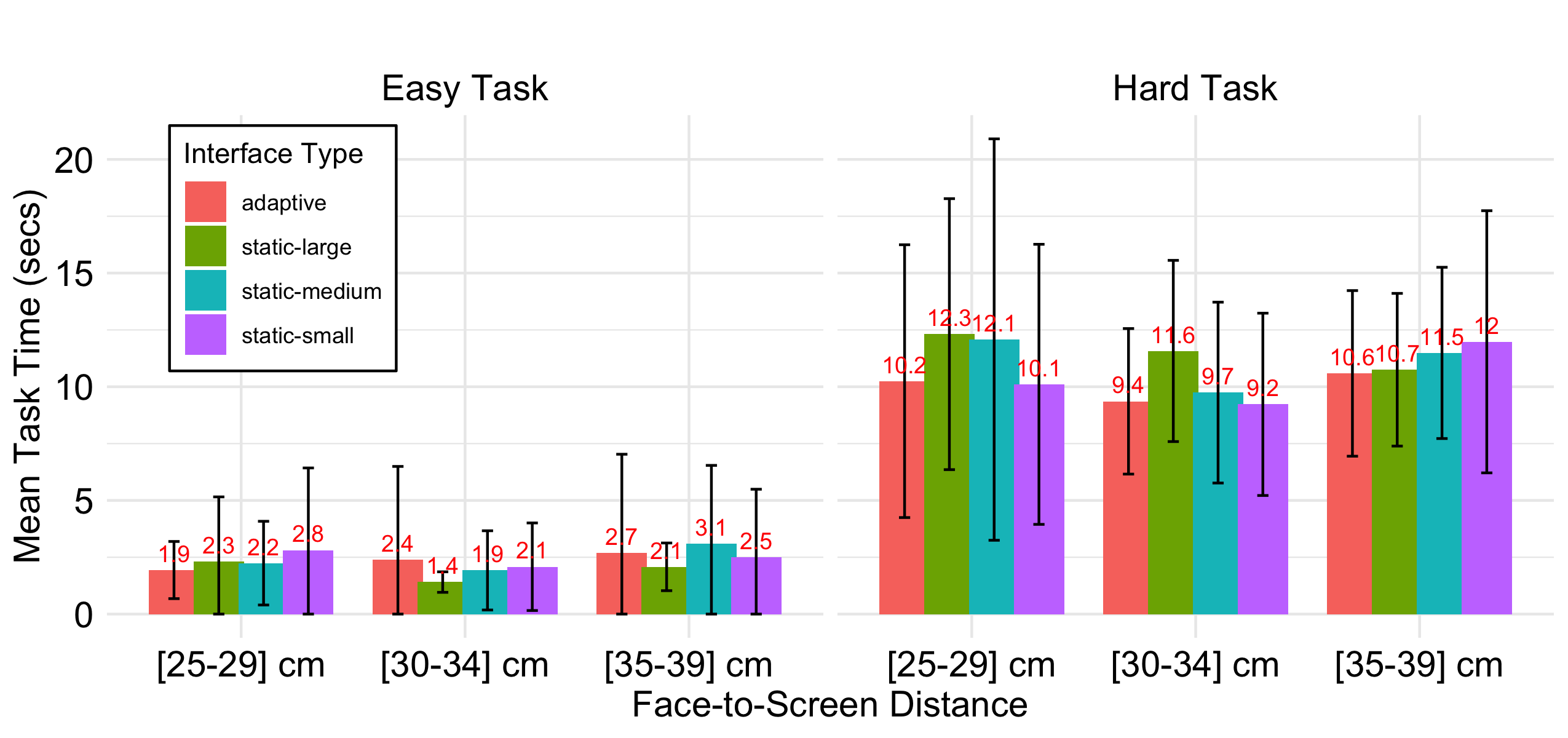}
    \label{fig:task_time_details}
  \caption{Mean task time by levels of \textit{Interface Types}, \textit{Face-to-Screen Distance} and \textit{Task Difficulty}. For the hard tasks, both \textit{Interface Type} and \textit{Face-to-Screen Distance} showed a significant main effect on task time. Error bars represent standard deviation. The \textit{adaptive} and \textit{static-small} resulted in significantly faster task time than the \textit{static-large} UI for the hard task. As expected, hard tasks took longer than easy ones across all conditions.}
  \Description{Two figures side by side, where the x-axis represents the face-to-screen distance and the y-axis shows the mean task time in seconds from 0 to 20. The left figure represents the easy task results, while the right figure shows the results for the hard task. All the data is represented in a bar chart with error bars representing the STD grouped by the face-to-screen distance for each of the four distances, 25-29 cm, 20-34 cm, and 35-39 cm. The results are represented by the interface type, namely, \textit{adaptive}, \textit{static-large}, \textit{static-medium}, and \textit{static-small}.}
  \label{fig:task_time}
\end{figure}

For statistical tests, we used repeated measures ANOVA with Aligned Rank Transform (ART) \cite{10.1145/1978942.1978963} applied in cases where data transformation was required, such as when the collected data was not normally distributed, as determined by the Shapiro-Wilk test. We reported post hoc pairwise comparisons with Bonferroni correction, using ART-C \cite{10.1145/3472749.3474784} if ART was conducted. Where there is an interaction effect, separate one-way repeated measures ANOVA tests were run too to distinguish the impact of each condition. Based on the study design and procedures, we collected measures from a total of 1728 trials (24 participants $\times$ 3 face-to-Screen distances $\times$ 4 interface types  $\times$ 2 task difficulties $\times$ 3 trials). We excluded data from two participants (P10 and P16) in Phase 1 because the tracking library failed to capture 38.45\% and 54.95\% of frames at the [25–29] cm distance, both more than two standard deviations above the sample mean ($M = 6.24$, $SD = 13.3$). 
Out of the remaining data (1584), data from 19 trials were excluded because participants self-reported not paying attention during the task, either when searching for the soundtrack or forgetting its title.

\subsection{Task Time}
Figure \ref{fig:task_time} shows the descriptive statistics.

For the easy tasks, there was a significant main effect of \textit{Face-to-Screen Distance}, F\textsubscript{2,231} = 3.28, $p<.05$ on task time. Post hoc analysis revealed that participants completed tasks significantly ($p<.05$) faster at a viewing distance of [30-34] cm ($M=1.95s$, $SD=1.23$) compared to [35-39] cm ($M=2.59s$, $SD=2.08$). The other pairs were not significantly different ($p>.05$).

For the hard tasks, there was a significant main effect of \textit{Interface Type}, F\textsubscript{3,231} = 3.29, $p<.05$ on the task time. Post hoc analysis showed that overall, the \textit{adaptive} ($M=10.1s$, $SD=4.42$) and \textit{static-small} ($M=10.4s$, $SD=5.43$) interfaces resulted in significantly ($p<.05$) faster task time compared to the \textit{static-large} interface ($M=11.5s$, $SD=4.55$). No significant differences were found between other pairs (see Table \ref{tab:task_time}). 
There was a significant main effect of \textit{Face-to-Screen Distance}, F\textsubscript{2,231} = 4.50, $p<.05$ on task time. Post hoc analysis showed that the farthest distance, i.e., [35-39] cm ($M=11.20$, $SD=4.21$), resulted in significantly ($p<.05$) slower task time compared to closer distances. The mean values were $11.19s$ at a distance of [25-29] cm ($SD=6.80$) and $9.98s$ at a distance of [30-34] cm ($SD=3.86$).

\revision{Overall, although the \textit{adaptive} interface required navigation similar to the \textit{static-large} UI at the farthest viewing distance (35-39 cm), both the \textit{static-small} and the \textit{adaptive} UIs resulted in significantly faster task time than the \textit{static-large} UI}.

\begin{table}[]
\caption{Task time (in seconds) for \textit{Interface Type} and \textit{Task Difficulty}. Standard deviations are in parentheses. For the hard tasks, the \textit{adaptive} and \textit{static-small} interfaces resulted in significantly faster task times compared to the \textit{static-large} interface.}
\label{tab:task_time}
\begin{tabular}{|l|c|c|}
\hline
\multicolumn{1}{|c|}{\textbf{Interface Type}} & \textbf{easy} & \textbf{hard} \\ \hline
adaptive                                      & 2.33 (3.49)   & 10.1 (4.42)   \\ \hline
static-large                                  & 1.93 (1.78)   & 11.5 (4.55)   \\ \hline
static-medium                                 & 2.42 (2.48)   & 11.1 (5.99)   \\ \hline
static-small                                  & 2.46 (2.90)   & 10.4 (5.43)   \\ \hline
\end{tabular}
\end{table}

\subsection{Navigation Time}
\begin{figure}[t]
    \centering
    \includegraphics[width=0.9\linewidth]{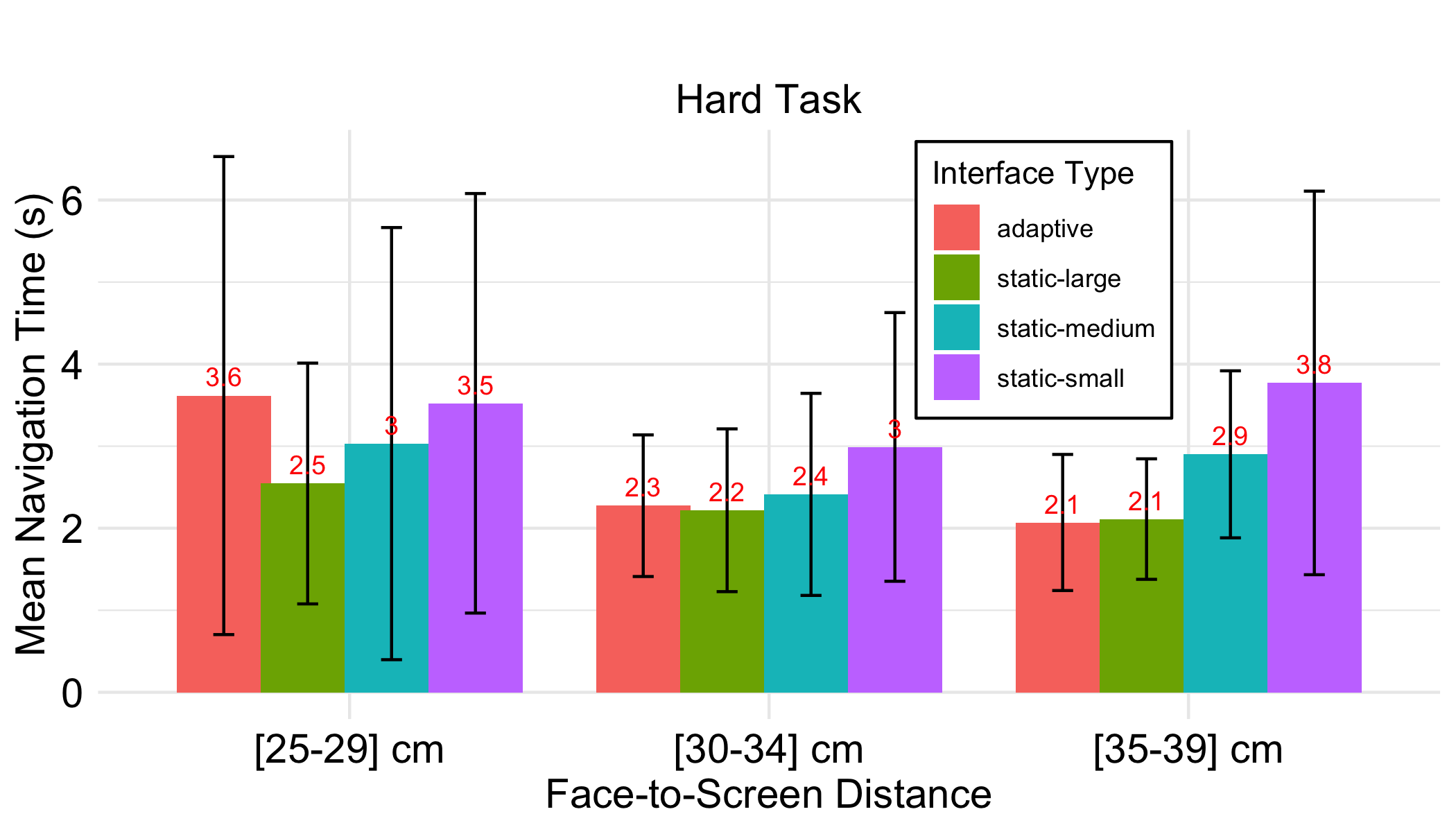}
    \caption{Mean navigation time by levels of \textit{Interface Type} and \textit{Face-to-Screen Distance}. At the farthest distance between the participants and the phone, the \textit{adaptive} and \textit{static-large} interfaces significantly reduced navigation time compared to the \textit{static-medium} and \textit{static-small} interfaces. Error bars represent standard deviations.} 
    \Description{Shows three groups of bar charts. The x-axis represents the face-to-screen distance, while the y-axis represents the mean navigation time in seconds for the hard task from 0 to 6. The figures show three groups of bar charts representing three face-to-screen distances: 25-29 cm, 20-34 cm, and 35-39 cm. The results are represented by the interface type, namely, \textit{adaptive}, \textit{static-large}, \textit{static-medium}, and \textit{static-small} .}
    \label{fig:navigation_time}
\end{figure}

 Figure \ref{fig:navigation_time} shows the descriptive statistics for the hard tasks, as the easy tasks did not require any navigation step.
 
We found a statistically significant interaction effect between \textit{Interface Type} and \textit{Face-to-Screen Distance} on navigation time for the hard tasks, F\textsubscript{6, 231} = 3.04, $p < .01$. Post hoc analysis showed that at the viewing distance of [35-39] cm, the \textit{adaptive} ($M=2.07s$, $SD=.83$) and the \textit{static-large} ($M=2.11s$, $SD=.73$) interfaces resulted in significantly ($p< .001$) faster navigation time compared to the \textit{static-medium} ($M=2.90s$, $SD=1.02$) and \textit{static-small} ($M=3.77s$, $SD=2.34$) interfaces. We found no significant differences in navigation time between interfaces at other viewing distances.

The results suggest that for the hard tasks, the \textit{adaptive} and the \textit{static-large} interfaces helped reduce the navigation time significantly at the viewing distance of [35-39]cm, compared to the other two interfaces.

\subsection{Task Error Counts}
\begin{figure}[!t]
    \centering
    \includegraphics[width=\linewidth]{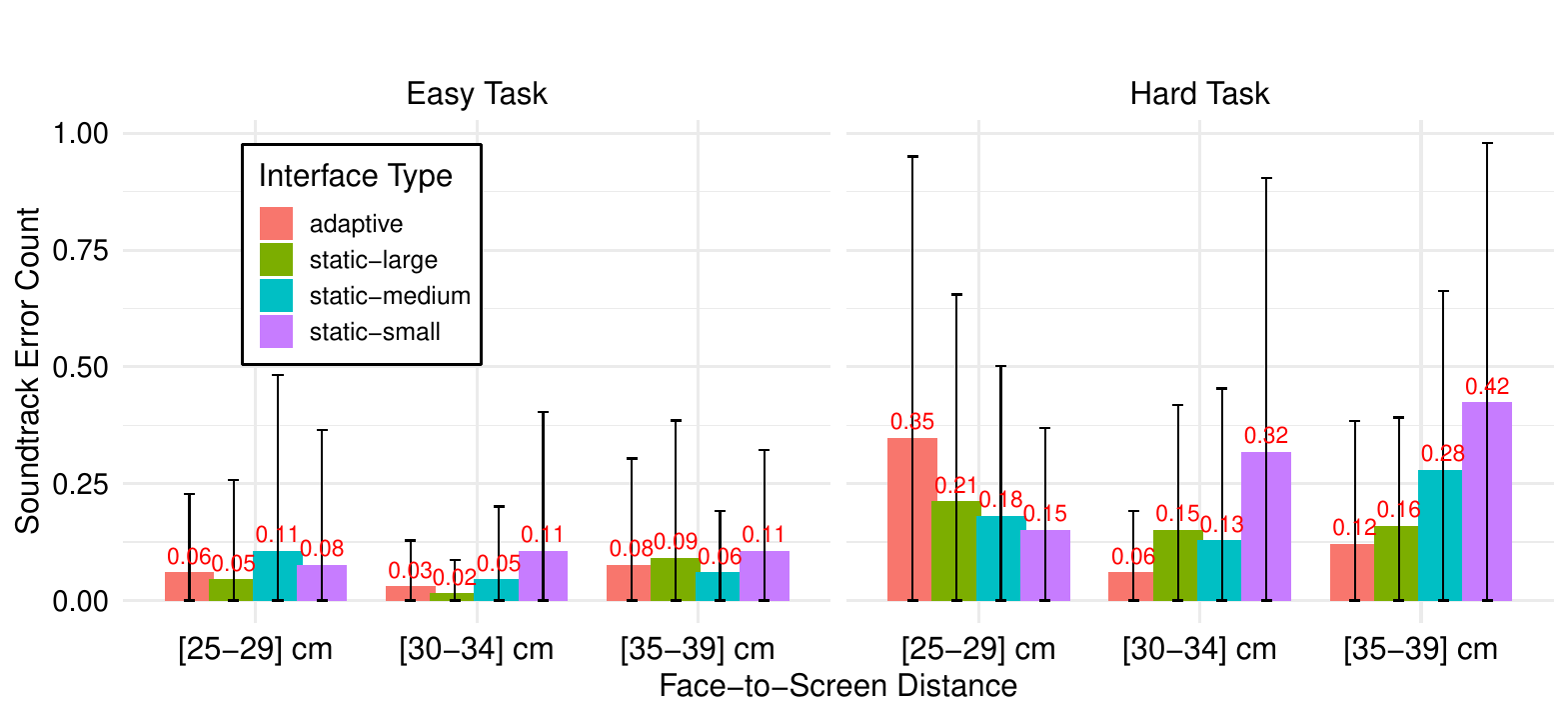}
    \caption{We counted the number of times participants played incorrect soundtracks before playing the correct one. While there were no significant differences between the interfaces on error in the easy task, the \textit{adaptive} UI resulted in significantly lower error compared to the \textit{static-small} UI at the farthest distance; i.e., [35-39] cm. Error bars represent the standard deviations.} 
    \Description{Two figures representing the soundtrack error count on the y-axis from 0 to 1, for the easy task on the left side and the hard task on the right side. The x-axis represents the face-to-screen distance for these groups 25-29 cm, 20-34 cm, and 35-39 cm. The results are represented by the interface type, namely, \textit{adaptive}, \textit{static-large}, \textit{static-medium}, and \textit{static-small} .}
    \label{fig:error_count_soundtracks}
\end{figure}

We considered two types of errors as defined in Section \ref{sec:measures}, the soundtrack and play/pause control errors. See Figure \ref{fig:error_count_soundtracks} for soundtrack errors descriptive statistics.

For the soundtrack errors, we counted the errors when playing the incorrect soundtrack only.
For the easy tasks, the main effect of \textit{Interface Type} or \textit{Face-to-Screen Distance} was not significant on error, nor was there an interaction between the two. 
For the hard tasks, there was a significant interaction effect for \textit{Interface Type}*\textit{Face-to-Screen Distance} on errors, F\textsubscript{6,231} = 2.50, $p<.05$. Pairwise comparisons revealed that the \textit{adaptive} interface ($M=.12$, $SD=.26$) resulted in significantly ($p<.05$) reduced errors compared to the \textit{static-small} interface ($M=.42$, $SD=.56$) at the distance of [35-39] cm. No significant differences were found between other pairs.

For the play/pause control errors, we counted the errors that occurred when the play/pause buttons were inadvertently activated while navigating the soundtrack list. We calculated this error only for the hard tasks, as the easy tasks did not require any navigation steps.
For the hard tasks, there was a significant main effect of \textit{Interface Type} on the errors, F\textsubscript{3,231}=$3.86$, $p<.05$. Post hoc analysis showed that the \textit{adaptive} ($M=.18$, $SD=.51$) and \textit{static-medium} ($M=.19$, $SD=.46$) interfaces caused significantly ($p<.05$) less errors compared to the \textit{static-small} ($M=.27$, $SD=.63$). There was also a significant main effect of \textit{Face-to-Screen Distance} on the error, F\textsubscript{2,231}=$6.42$, $p<.01$. The pairwise comparison revealed that navigating the soundtrack at the distances of [25-29] cm ($M=.1$, $SD=.3$) and [30-34] cm ($M=.22$, $SD=.79$) resulted in significantly ($p<.05$) fewer errors than at [35-39] cm ($M=.40$, $SD=.86$). 

Overall, the \textit{adaptive} interface helped reduce errors when playing soundtracks at increased distances, i.e., [35-39] cm, compared to the \textit{static-small} interface. Both the \textit{adaptive} and \textit{static-medium} interfaces reduced the likelihood of accidentally activating the play/pause button when navigating the list of soundtracks, compared to the \textit{static-small}.

\subsection{Timeouts}
We counted the number of times participants failed to complete the task within the given time, resulting in a timeout (see Table \ref{tab:timeout}). Out of the 1565 trials, 19 trials resulted in a timeout. We found that 12 out of the 19 timeouts occurred when the participants were performing the hard task at the closest distance to the phone, i.e., [25-29] cm. No timeouts were registered for the easy task. Also, over half of the timeouts observed during the trials were attributed to Participants 2 and 4, with each contributing five timeouts. Out of the 22 participants, 15 did not experience any timeouts.

\begin{table}[]
\centering
\caption{In our experiment, all the timeouts occurred when the participants were doing the hard tasks. The easy tasks did not result in any timeout.}
\label{tab:timeout}
\begin{tabular}{l|lll}

                        & \multicolumn{3}{c}{\textbf{Face-to-Screen Distance}}                                                                  \\ \hline
\textbf{Interface Type} & \multicolumn{1}{l|}{\textbf{{[}25-29 {]} cm}} & \multicolumn{1}{l|}{\textbf{{[}30-34{]} cm}} & \textbf{{[}35-39{]} cm} \\ \hline
adaptive                & \multicolumn{1}{l|}{1}                        & \multicolumn{1}{l|}{0}                       & 1                       \\ \hline
static-large            & \multicolumn{1}{l|}{3}                        & \multicolumn{1}{l|}{0}                       & 1                       \\ \hline
static-medium           & \multicolumn{1}{l|}{4}                        & \multicolumn{1}{l|}{4}                       & 0                       \\ \hline
static-small            & \multicolumn{1}{l|}{4}                        & \multicolumn{1}{l|}{1}                       & 0                       \\ \hline
\end{tabular}
\end{table}

\subsection{Perceived Workload}
Using NASA-TLX \cite{doi:10.1177/154193120605000909}, we collected responses from participants to rate their perceived workload while using the interfaces. The overall mean scores, out of 100 \cite{10.1145/3544548.3580871}, were calculated across all six workload dimensions for each interface. 
We found a significant main effect of \textit{Interface Type}, F\textsubscript{3,63}= 4.08, $p<.05$. Participants perceived \textit{static-medium} ($M=25.9$, $SD=16.5$) and \textit{static-large} ($M=26.1$, $SD=18.2$) to cause significantly ($p<.05$) lower workloads compared to \textit{static-small} ($M=33.8$, $SD=21.6$). We found no evidence for significant differences between the \textit{adaptive} interface ($M=28.3$, $SD= 20.1$) and the other interfaces in terms of workload.

\begin{table*}[!t]
\centering
\caption{Participant reported on usability aspects of the four \textit{Interface Types} on a 5-point Likert scale (1=Strongly Disagree;5=Strongly Agree). The table shows the median reported ratings and Friedman test results. The IQR values are given in brackets. The reported results indicate that we have no evidence of an additional penalty due to adaptation on usability compared to static interfaces.} 
\label{tab:likert}
\resizebox{\textwidth}{!}{%
\begin{tabular}{lllll|ll}

\multicolumn{1}{c}{\textbf{Usability aspect}}               & \multicolumn{4}{c}{\textbf{Interface Type}}                                                                                                              & \multicolumn{2}{c}{\textbf{Friedman}}   \\ \hline
\multicolumn{1}{c|}{~}
                                                            & \multicolumn{1}{l|}{\textbf{adaptive}} & \multicolumn{1}{l|}{\textbf{static-large}} & \multicolumn{1}{l|}{\textbf{static-medium}} & {\textbf{static-small}} & \multicolumn{1}{l|}{$\chi^{2}(3)$} & {\textbf{\textit{p}}} \\ \cline{2-7} 
\multicolumn{1}{l|}{\textbf{Ease of Use (\revision{when viewing distance varies})}}  & \multicolumn{1}{l|}{4 (1)}             & \multicolumn{1}{l|}{4 (1.75)}              & \multicolumn{1}{l|}{4 (1)}                  & 4 (1)                 & \multicolumn{1}{l|}{2.37}   & 0.50       \\ \hline
\multicolumn{1}{l|}{\textbf{Learnability}}                 & \multicolumn{1}{l|}{5 (1)}             & \multicolumn{1}{l|}{5 (1)}                 & \multicolumn{1}{l|}{5 (1)}                  & 5 (1)                 & \multicolumn{1}{l|}{4.91}   & 0.18       \\ \hline
\multicolumn{1}{l|}{\textbf{Low attention demand (\revision{when viewing distance varies})}}         & \multicolumn{1}{l|}{3 (2)}             & \multicolumn{1}{l|}{3.5 (2)}               & \multicolumn{1}{l|}{3 (3)}                  & 3 (2)                 & \multicolumn{1}{l|}{2.74}   & 0.43       \\ \hline
\multicolumn{1}{l|}{\textbf{High eye fatigue (\revision{when viewing distance varies})}}  & \multicolumn{1}{l|}{4 (1)}             & \multicolumn{1}{l|}{3 (2)}                 & \multicolumn{1}{l|}{3 (1)}                  & 3 (1.75)              & \multicolumn{1}{l|}{2.94}   & 0.40       \\ \hline
\multicolumn{1}{l|}{\textbf{\revision{Perceived Accuracy} (\revision{when viewing distance varies})}}     & \multicolumn{1}{l|}{4 (1)}             & \multicolumn{1}{l|}{4 (1)}                 & \multicolumn{1}{l|}{4 (1)}                  & 4 (1.5)               & \multicolumn{1}{l|}{3.37}   & 0.34       \\ \hline
\multicolumn{1}{l|}{\textbf{\revision{Perceived Speed} (\revision{when viewing distance varies})}}        & \multicolumn{1}{l|}{4 (2)}             & \multicolumn{1}{l|}{4 (1.75)}              & \multicolumn{1}{l|}{4 (1.75)}               & 4 (2)                 & \multicolumn{1}{l|}{2.16}   & 0.54       \\ \hline
\multicolumn{1}{l|}{\textbf{Efficiency (Easy task)}}       & \multicolumn{1}{l|}{4.5 (1)}           & \multicolumn{1}{l|}{5 (1)}                 & \multicolumn{1}{l|}{5 (1)}                  & 5 (1)                 & \multicolumn{1}{l|}{3.00}   & 0.39       \\ \hline
\multicolumn{1}{l|}{\textbf{Efficiency (Hard task)}}       & \multicolumn{1}{l|}{4 (1.75)}          & \multicolumn{1}{l|}{4 (1.75)}              & \multicolumn{1}{l|}{4 (1.75)}               & 4 (1)                 & \multicolumn{1}{l|}{3.57}   & 0.31       \\ \hline
\multicolumn{1}{l|}{\textbf{Enjoyability (\revision{when viewing distance varies})}} & \multicolumn{1}{l|}{4 (1)}             & \multicolumn{1}{l|}{4 (2)}                 & \multicolumn{1}{l|}{4 (2)}                  & 4 (2)                 & \multicolumn{1}{l|}{4.42}   & 0.22       \\ \hline
\multicolumn{1}{l|}{\textbf{Comfort (\revision{when viewing distance varies})}}      & \multicolumn{1}{l|}{4 (1.75)}          & \multicolumn{1}{l|}{4 (1)}                 & \multicolumn{1}{l|}{4 (2)}                  & 3 (2.75)              & \multicolumn{1}{l|}{5.38}   & 0.15       \\ \hline
\multicolumn{1}{l|}{\textbf{Daily use intent}}             & \multicolumn{1}{l|}{4 (2)}             & \multicolumn{1}{l|}{3.5 (1)}               & \multicolumn{1}{l|}{4 (2)}                  & 4 (2)                 & \multicolumn{1}{l|}{5.36}   & 0.15       \\ \hline
\end{tabular}%
}
\end{table*}

\subsection{Subjective Feedback}
We asked the participants to rate their experience with each \textit{Interface Type} using 5-point Likert scale questions, where 1 indicated \emph{\say{Strongly Disagree}} and 5 indicated \emph{\say{Strongly Agree}}. \revision{We structured the questions as statements. These included: \textit{"When the distance changes, the interface is \{ easy to use, easy to learn, does not require much attention, tiring to the eyes, enjoyable to use, comfortable to use \}"}. We also included the following: \textit{"When the distance changes, the interface enables me to accomplish the task \{accurately, quickly, efficiently when the song is on the current page [Easy Task], efficiently when the song is on subsequent pages [Hard Task]"}. We did not use the labels Easy and Hard in the actual statements to avoid any unintended interpretations these terms might convey; they are provided here solely for clarification. In addition, we included the statement \textit{"I will use this interface daily to accomplish tasks".}} We used the Friedman test to check for significance, and the Wilcoxon signed-rank test for pairwise comparisons with the Bonferroni correction. Table \ref{tab:likert} presents a summary of the participants' ratings on various aspects, along with the Friedman test results. Overall, we found no evidence that the \textit{adaptive} UI was perceived differently or worse than the static interfaces, as usability medians did not differ significantly across interfaces.

\begin{figure}[!t]
    \centering
    \includegraphics[width=0.99\linewidth]{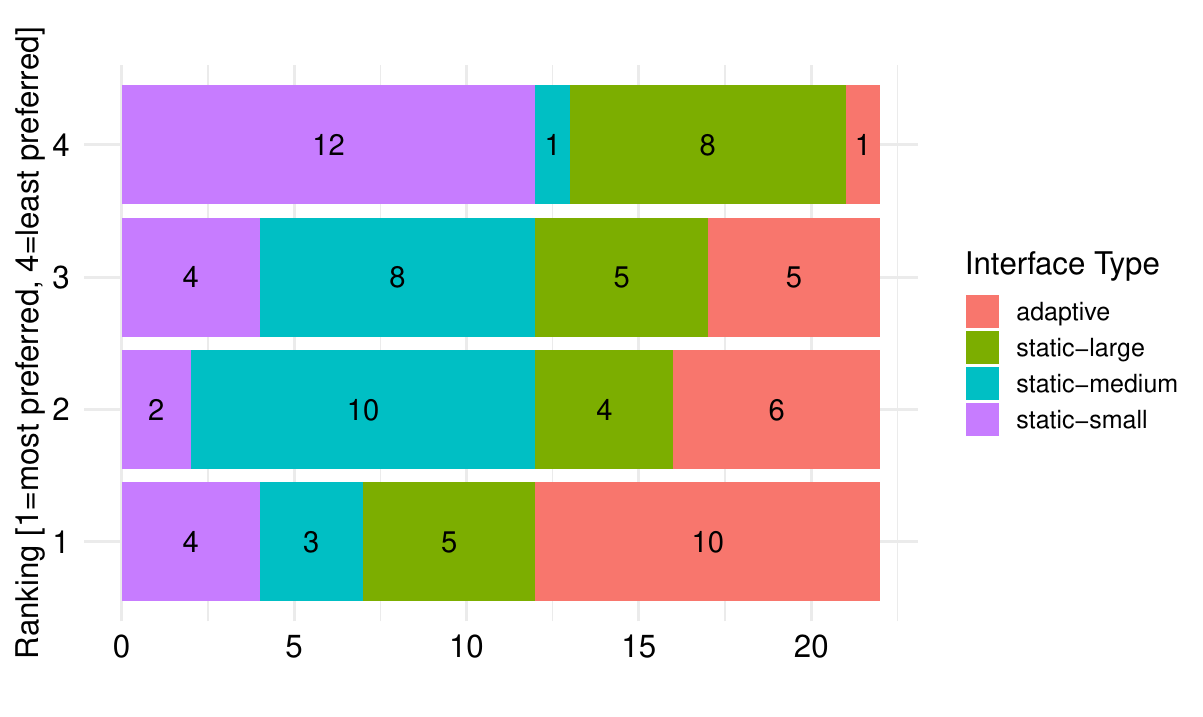}
    \Description{Stack chart representing the ranking of the interface type from 0 to 24 (number of participants) for the four interface types, \textit{adaptive}, \textit{static-large}, \textit{static-medium}, and \textit{static-small} .}
    \caption{Participants ranked their preference for the \textit{Interface Types} after experiencing all the different interfaces in phase 1 of the experiment (1=Most preferred; 4=Least preferred). While the \textit{adaptive} interface was ranked the most or second-most preferred interface by 16 participants, the \textit{static-small} was the least or second-least preferred interface by 16 participants. The \textit{static-medium} interface was the second most preferred interface type.}
    \label{fig:ranking}
\end{figure}

\subsection{Interface Type Ranking}
\revision{Upon completing the tasks in Phase 1,} we asked the participants to express their preference for the different UIs, assigning rankings from 1 to 4, (1= Most preferred; 4= Least preferred) (see Figure \ref{fig:ranking}).
To compute the overall ranking score for each interface, we applied a weighted ranking method on participants’ preferences. For each interface, we counted the number of times it was assigned to each rank. These counts were then weighted to reflect the relative importance of each rank. A rank of 1 was assigned a weight of 4, rank 2 a weight of 3, rank 3 a weight of 2, and rank 4 a weight of 1. The count for each rank was multiplied by its corresponding weight, and the resulting values were summed to produce a total weighted score for each interface.
Among the 22 participants, the \textit{adaptive} UI received the highest preference ($score= 69$), with 10 ranking it most preferred and 6 second ($Med=2$, $Q1=1$, $Q3=2.75$). 
Many found the \textit{adaptive} UI \emph{easy to use,} \emph{convenient,} and \emph{easiest to complete.} 
Several participants appreciated its adaptive design, which made the UI more comfortable and accurate, \emph{\say{[the \textit{adaptive} UI] allowed for better recognition of where my eyes were, and for me to choose the correct options more efficiently}} (P21), \emph{\say{adapting to the distance is a good idea for more accurate results}} (P7). P11 reported that having too many items made it hard to focus, whereas too few items required more navigation, suggesting that an \textit{adaptive} UI could help balance visual load and efficiency.
While the \textit{static-medium} UI was the second most preferred ($score=59$, $Med=2$, $Q1=2$, $Q3=3$), the \textit{static-large} UI was the second least preferred ($score=50$, $Med=3$, $Q1=2$, $Q3=4$). 
For the \textit{static-medium} UI, few participants appreciated that they did not need to change too many pages, found the UI visually comfortable, and noted that a familiar layout made it easier to navigate.
However, two participants raised concerns. One found that the \textit{static-medium} took longer to recognise the eyes, and another felt that the smaller navigation buttons detracted from overall usability.
\revision{On the other hand, several participants found the \textit{static-large} UI generally easy to use, visually clear, and easy to follow, and they reported improved perceived accuracy,} with one highlighted the usefulness of larger navigation buttons. However, a few participants disliked the frequent need to navigate pages, \emph{\say{I did not like [the static-large], because in all cases I needed to change pages more often}} (P5). Two participants also felt that elements were too large, leading to minor visual fatigue.
The \textit{static-small} UI was ranked the least preferred by 16 participants ($score=42$), who ranked it either at the bottom or second to last ($Med=4$, $Q1=2.25$, $Q3=4$). While a few participants appreciated the ability to view more soundtracks on a single screen with the \textit{static-small} UI, which reduced the need for extra navigation, some participants found that too many items at once made it harder to focus, and the smaller elements occasionally led to tracking issues, degrading usability.

\subsection{Participants Experience \revision{ with Natural Use of GAUI (Phase 2)}} \label{sec:phase2}
In the second phase of the experiment, we allowed participants to interact with the \revision{\methodname-based media player naturally}, in six different postures: while walking, standing, slouching, sitting (hands-free), sitting (desk), and also sitting (chair). \revision{In this phase, we did not control or distribute the viewing distance, allowing participants to interact freely with the application. This phase enabled us to capture the participants' perception of using the novel \methodname naturally, given the variation in postures, which inherently vary distances \cite{10.1145/3025453.3025794}.}
In each posture, the application was used for two minutes. 
During the interaction, we logged data, including face distance from the screen, while also measuring participants' subjective experiences using experience-sampling questions presented to them in each posture after an adaptation was detected. Further details of the method used are provided in Section \ref{sec:measures}. 
Due to technical issues, data for one posture from P17 were missing. We collected responses only when a UI adaptation took place (the event).

\begin{figure}[!t]
    \centering
    \includegraphics[width=0.9\linewidth]{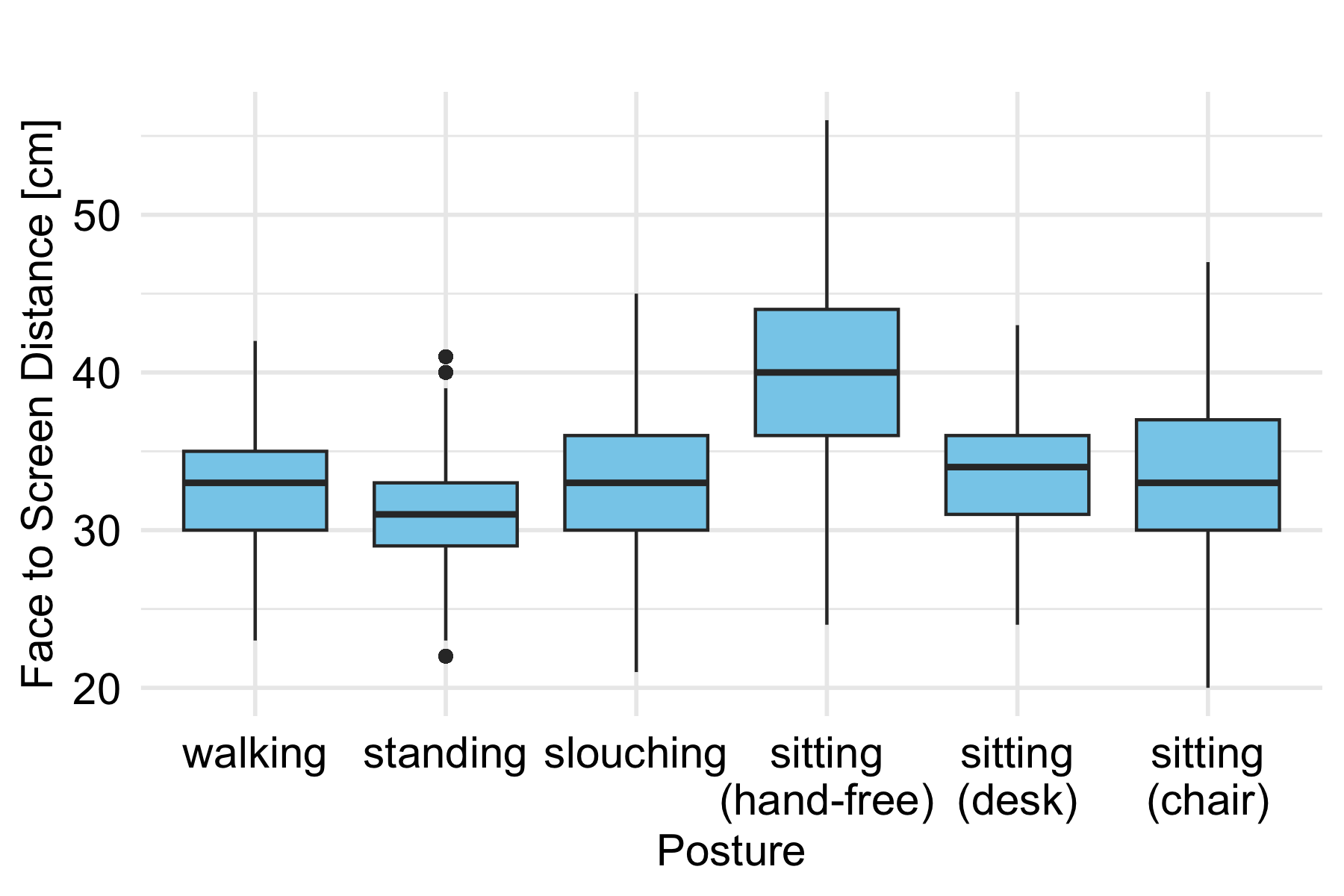}
    \caption{The range of the measured face-to-screen distances during the second phase of the experiment across six different postures. The thick black lines indicate the median, while the boxes denote the 25th and 75th percentiles. Participants showed similar viewing distances for each holding posture as in prior work \cite{10.1145/3025453.3025794}.
    } 
    \Description{A box plot showing the face-to-screen distance in cm on the y-axis from 0 to 60 cm and the posture on the x-axis, namely walking, standing, slouching, sitting (hand free), sitting (desk), and sitting (chair).}
    \label{fig:face_to_Screen}
\end{figure}

\subsubsection{Viewing Distances Per Posture}
We measured the distances at which participants held the phone while interacting with the \revision{\methodname-based media player across the six postures (see Figure \ref{fig:face_to_Screen})}. When the phone is used hands-free, the distance is much greater ($Med=41cm$, $Q1=36cm$, $Q3=44cm$) compared to other postures. \revision{Overall, the average viewing distance across all other postures ranged between 30 and 40 cm, indicating that participants held the phone at viewing distances consistent with prior work \cite{10.1145/3025453.3025794}}.

\subsubsection{Viewing Distance Changes Within Each Posture}
The mean of the number of times participants switched the viewing distances in each posture is as follows, with the standard deviation in brackets: 

\begin{table}[h]
    \centering
    \rowcolors{2}{white}{white}
    \resizebox{\columnwidth}{!}{%
    \begin{tabular}{c c c c c c}
    \rowcolor{gray!20}
    \multicolumn{6}{c}{\textbf{Mean Viewing Distance Switches in Each Posture}} \\
        \textbf{Walking} & Standing & Slouching & Sitting (hands-free) & Sitting (desk) & Sitting (chair) \\
        \textbf{22.5 (34.8)}     & 9.58 (12.1)  & 11.3 (15.4)    &  5.46 (9.43)    & 9.33 (15.3)     & 11.3 (17.1)     \\
    \end{tabular}}
\end{table}

We observed that few participants switched distances much more than the mean average ($M= 11.54, SD= 10.16$), such as P5 ($M= 43.2, SD= 26.9$), P8 ($M= 29.7, SD= 69.2$), and P23 ($M= 43.5, SD= 11.6$), while some participants did not switch distance at all in each posture, maintaining a static distance between them and the phone:  

\begin{table}[h]
    \centering
    \rowcolors{2}{white}{white}
    \resizebox{\columnwidth}{!}{%
    \begin{tabular}{c c c c c c}
    \rowcolor{gray!20}
    \multicolumn{6}{c}{\textbf{"No Viewing Distance Switching" – Number of Participants}} \\
        Walking & Standing & Slouching & \textbf{Sitting (hands-free)} & Sitting (desk) & Sitting (chair) \\
        3     & 7  & 3    & 12    & 6     & 5     \\
    \end{tabular}}
\end{table}

\subsubsection{Awareness and Expectation}
We assessed whether participants were aware of the adaptation and whether the UI changes aligned with their expectations. To measure this, we asked two questions. The first was: \emph{\say{Did you notice the user interface adjusting size just now?}}, with response options: \emph{\say{Yes, Unsure, and No}}. The second question was: \emph{\say{The user interface resized the way I expected}} (1=Strongly Disagree; 5=Strongly Agree). The majority of participants reported noticing the UI changes. Their detailed responses are in the table below: 

\begin{table}[h]
    \centering
    \rowcolors{2}{white}{white}
    \resizebox{\columnwidth}{!}{%
    \begin{tabular}{c c c c c c}
    \rowcolor{gray!20}
    \multicolumn{6}{c}{\parbox{.5\textwidth}{\centering \textbf{"Did you notice the user interface adjusting size just now?"}\\\textbf{(Yes, No, Unsure)}}} \\
        Walking & Standing & Slouching & Sitting (hands-free) & Sitting (desk) & Sitting (chair) \\
        \textbf{Yes (90.5\%)}  & \textbf{Yes (93.3\%)}  & \textbf{Yes (80\%)}    & \textbf{Yes (83.3\%)}    & \textbf{Yes (94.4\%)}     & \textbf{Yes (72.2\%)}     \\

         Unsure (0\%)      & Unsure (6.67\%)  & Unsure (15\%)    & Unsure (8.33\%)    & Unsure (5.56\%)     & Unsure (16.7\%)  \\ 
        No (9.52\%)      & No (0\%)  & No (5\%)    & No (8.33\%)    & No (0\%)     & No (11.1\%)     \\

    \end{tabular}}
\end{table}

Regarding the participants' expectations, while they generally reported that the interface resized as expected, walking received the lowest rating ($Med=3$, $Q1=3$, $Q3=4$). The table below reports the median, first and third quartiles based on participants' ratings of expectation in each posture. The posture with the lowest rating is highlighted in bold.

\begin{table}[h]
    \centering
    \rowcolors{2}{white}{white}
    \resizebox{\columnwidth}{!}{%
    \begin{tabular}{c c c c c c}
    \rowcolor{gray!20}
    \multicolumn{6}{c}{\parbox{.5\textwidth}{\centering \textbf{"The user interface resized the way I expected?"}\\\textbf{(1=Strong disagree; 5= Strongly agree)}}} \\
        \textbf{Walking} & Standing & Slouching & Sitting (hands-free) & Sitting (desk) & Sitting (chair) \\
        \textbf{3 ($3-4$)}  & 4 ($3.5-4.5$)   & 4 ($3-4.25$)     & 4 ($3.75-5$)     & 4 ($3.25-4.75$)     & 3.5 ($3-4.75$)   
          
    \end{tabular}}
\end{table}

\subsubsection{Influence Over UI Adaptation}
To assess whether participants felt they had any influence over the UI changes in each posture, we asked: \emph{\say{How much influence did you feel you had over the changes in the user interface?}} using a 5-point Likert scale (1= Not at all; 5= Completely). Across all postures, the median response was 3, suggesting that participants \emph{\say{somewhat}} felt in control of the UI changes. Postures such as sitting (desk) ($Med=3.5$, $Q1=3$, $Q3=4$), sitting (hands-free) ($Med=4$, $Q1=2.75$, $Q3=4$), and slouching ($Med=3$, $Q1=3$, $Q3=5$) reflected a slightly stronger sense of influence. The median values, along with their first and third quartiles, are presented below. Postures with a lower sense of influence are highlighted in bold. 

\begin{table}[h]
    \centering
    \rowcolors{2}{white}{white}
    \resizebox{\columnwidth}{!}{%
    \begin{tabular}{c c c c c c}
    \rowcolor{gray!20}
    \multicolumn{6}{c}{\parbox{.5\textwidth}{\centering \textbf{"How much influence did you feel you had over the changes in the user interface?"}\\\textbf{(1=Not at all; 5= Completely)}}}
     \\
    \textbf{Walking} & \textbf{Standing} & Slouching & Sitting (hands-free) & Sitting (desk) & \textbf{Sitting (chair)} \\
        \textbf{3 ($2-4$)}  & \textbf{3 ($2-3.5$)}   & 3 ($3-5$)     & 4 ($2.75-4$)     & 3.5 ($3-4$)     & \textbf{3 ($1.25-4$)}   
          
    \end{tabular}}
\end{table}

\subsubsection{Helpfulness and Satisfaction}
To assess perceived helpfulness and satisfaction with the user interface adaptation \revision{when interacting with \methodname-based media player}, participants responded to two Likert-scale questions: \emph{\say{I found this user interface change helpful}} and \emph{\say{I am satisfied with how the user interface adapted,}} (1=Strongly Disagree; 5=Strongly Agree). While across most postures, the median response for both questions was consistently 4, indicating generally positive perceptions, walking posture had the lowest rating for usefulness of adaptation ($Med=3$, $Q1=2$, $Q3=4$) and a first quartile of 2 for satisfaction ($Med=4$, $Q1=2$, $Q3=4$), suggesting that participants found the adaptation less helpful and less satisfying when walking. The table below reports the median, first and third quartiles based on participants' ratings. Postures with the lowest ratings are highlighted in bold.

\begin{table}[H]
    \centering
    \rowcolors{2}{white}{white}
    \resizebox{\columnwidth}{!}{%
    \begin{tabular}{c c c c c c}
    \rowcolor{gray!20}
    \multicolumn{6}{c}{\parbox{.5\textwidth}{\centering \textbf{"I found this user interface change helpful."}\\\textbf{(1=Strong disagree; 5= Strongly agree)}}} \\
        \textbf{Walking} & Standing & Slouching & Sitting (hands-free) & Sitting (desk) & Sitting (chair) \\
        \textbf{3 ($2-4$)}  & 4 ($3-4$)   & 4 ($3-4$)     & 4 ($3-5$)     & 4 ($3-4$)     & 4 ($3-4$)  \\
    \rowcolor{gray!20}
    \multicolumn{6}{c}{\parbox{.5\textwidth}{\centering \textbf{"I am satisfied with how the user interface adapted."}\\\textbf{(1=Strong disagree; 5= Strongly agree)}}} \\
        \textbf{Walking} & Standing & Slouching & Sitting (hands-free) & Sitting (desk) & Sitting (chair) \\
        \textbf{4 ($2-4$)}  & 4 ($3-4$)   & 4 ($3-4$)     & 4 ($3-4.25$)     & 4 ($4-4$)     & 4 ($3-4$)  
          
    \end{tabular}}
\end{table}

\subsubsection{Preference for Control}
To understand participants' preferences regarding control over UI changes when using \methodname, we asked whether they would rather adjust the user interface manually instead of relying on automatic adaptation, with response options \emph{\say{Yes, Unsure, and No}}. 
Across all postures, the percentage of \emph{\say{Yes}} responses ranged from 33.3\% to 53.3\%, peaking in walking and standing postures. The percentage of \emph{\say{No}} responses ranged from 16.7\% to 35\%, suggesting that in some cases, participants were content with the UI changes. Sitting (chair) resulted in the highest percentage of \emph{\say{Unsure}} responses, reflecting uncertainty about the adaptation behaviour. 
Overall, postures associated with higher bodily motion, such as walking, inclined participants to desire greater control over the changes. The detailed responses are in the table below.

\begin{table}[H]
    \centering
    \rowcolors{2}{white}{white}
    \resizebox{\columnwidth}{!}{%
    \begin{tabular}{c c c c c c}
    \rowcolor{gray!20}
   \multicolumn{6}{c}{\parbox{.5\textwidth}{\centering \textbf{"Would you rather adjust the user interface size manually?"}\\\textbf{(Yes, No, Unsure)}}} \\
        Walking & Standing & Slouching & Sitting (hands-free) & Sitting (desk) & Sitting (chair) \\
        \textbf{Yes (52.4\%)}  & \textbf{Yes (53.3\%)}  & Yes (35\%)    & Yes (41.7\%)    & Yes (33.3\%)     & Yes (33.3\%)     \\

         Unsure (23.8\%)      & Unsure (13.3\%)  & Unsure (30\%)    & Unsure (33.3\%)    & Unsure (33.3\%)     & \textbf{Unsure (50\%)}  \\ 
        No (23.8\%)      & No (33.3\%)  & No (35\%)    & No (25\%)    & No (33.3\%)     & No (16.7\%)     \\

    \end{tabular}}
\end{table}

\subsubsection{Adaptation Frequency/Sensitivity}
\revision{To evaluate \methodname's responsiveness and whether the frequency of its distance-based size changes felt appropriate}, we asked participants: \emph{\say{How do you feel about the frequency at which the user interface changed size?}}, with response options: \emph{\say{Just right, Not responsive enough, and Too sensitive}}. Across all postures, most participants felt the adaptation frequency was \textit{\say{just right}}. However, $38.1\%$ rated it as most sensitive while walking, suggesting they experienced more frequent or abrupt UI changes, likely due to increased bodily motion. 
The proportion of participants who felt that the UI was \emph{\say{not responsive enough}} was low across postures. These results suggest that while the adaptation frequency was broadly acceptable, certain postures, such as walking, may require a customised adaptation threshold to account for the dynamic environment. The detailed responses are provided in the table below.

\begin{table}[H]
    \centering
    \rowcolors{2}{white}{white}
    \resizebox{\columnwidth}{!}{%
    \begin{tabular}{c c c c c c}
    \rowcolor{gray!20}
    \multicolumn{6}{c}{\parbox{.6\textwidth}{\centering \textbf{"How do you feel about the frequency at which the user interface changed size?"}\\\textbf{(Just right (JR), Not responsive enough (NRE), Too sensitive (TS))}}} \\
        Walking & Standing & Slouching & Sitting (hands-free) & Sitting (desk) & Sitting (chair) \\
        \textbf{ JR (52.4\%)}  &  JR (80\%)  &  JR (80\%)    &  JR (75\%)    &  JR (66.7\%)     &  JR (75\%)     \\

         NRE  (9.52\%)      & NRE  (6.67\%)  & NRE  (5\%)    & NRE  (8.33\%)    & NRE  (11.1\%)     & NRE  (12.5\%)  \\ 
        \textbf{TS (38.1\%)}      & TS (13.3\%)  & TS (15\%)    & TS (16.7\%)    & TS (22.2\%)     & TS (12.5\%)     \\

    \end{tabular}}
\end{table}

\subsection{Qualitative Feedback}
We employed inductive thematic analysis \cite{braun2006using, braun2013successful} to analyse the exit interview data collected during the second phase of the experiment \ref{sec:phase2}. The leading researcher transcribed the interviews and reviewed the recordings to correct any errors in the auto-generated transcripts. The leading researcher and another researcher independently coded data from six participants using the online qualitative analysis platform QCAmap \cite{qcamap}, and reached agreement on the coding scheme. Following this, the lead researcher completed coding the remaining data, developed categories, and generated the final themes. The codes used were words or short phrases that captured key ideas expressed in the data.

\subsubsection{Positive and Negative Experiences with \methodname}
\revision{
As the participants experienced the \methodname-based media player in various postures, they were asked about their perception of it. Out of 24 participants, 15 expressed a positive experience using favourable language ($n=9$) (e.g., liking the UI and finding it useful), highlighting the functional benefits ($n=6$) (e.g., helpful, practical, improving the experience), and one noted that it was easy to learn. 
Participants also emphasised comfort and ergonomics ($n=2$), ease of use and control ($n=2$), and its innovative or enjoyable nature ($n=4$). 
P21 mentioned, \emph{\say{it was more comfortable to my eyes when the size changed, because if it maintains a single size, I feel like it would tire my eyes a bit}}. 
While very few participants were neutral about \methodname ($n=2$), seven did not enjoy it. They described the adaptation as making them feel lack of control ($n=1$), being difficult to use ($n=1$), or causing confusion or unintuitiveness ($n=5$), or initial confusion ($n=3$), \emph{\say{at first it was confusing. But, then, I got pretty used to it}} (P17).}

\subsubsection{Posture Matters: Positive in Static Postures and Negative in Walking}
\revision{Participants' perceptions of \methodname varied across postures.
Of the 24 participants, 15 reported positive experiences in static postures, describing it as comfortable ($n=2$), worked as expected ($n=2$), helpful in various static postures ($n=11$), accurate and efficient ($n=1$), well understood ($n=1$), and generally positive overall ($n=5$).
When walking, 13 participants reported a less favourable experience with \textit{adaptive} UI. They described it as contributing to cognitive overload ($n=7$), not helpful ($n=2$), difficult to use ($n=6$), less accurate ($n=3$), or simply disliked it ($n=1$) while walking. Several participants ($n=5$) felt that when walking, the UI changed too frequently due to movement; P12 commented, \emph{\say{it changed a lot. Even when I tried to hold it at one distance, it still changed}}. 
In contrast, one participant felt the \textit{adaptive} UI was better to use when standing and walking, and several perceived it positively while walking ($n=3$), describing it as helpful ($n=2$), user-friendly ($n=1$), adjusted to movement for ease of use ($n=1$), and improved accuracy ($n=1$). As P20 mentioned, \emph{\say{when I am walking … I usually have the phone further away … so having larger fields was helpful}}.
Two participants viewed the \textit{adaptive} UI negatively when sitting (hands-free) but preferred it when handheld ($n=1$). Another found it difficult to use when standing, and one reported challenges in most postures except sitting (chair).}

\subsubsection{Reflections on \methodname: Insights from Participants' Feedback}
\revision{Participants appreciated the benefits of switching to the \textit{static-large} UI at distances greater than 35 cm ($n=9$). 
While three responded positively to switching the UI to the \textit{static-small} at closer distances below 30 cm, seven were negative about it, \emph{\say{[it was] difficult to focus on a particular icon when you are too close. I think there needs to be a minimum size so they do not get too small}} (P6). 
While some participants appreciated the functional benefits of \methodname ($n=3$), one found the \textit{adaptive} UI helpful but non-essential and said they would not pay extra for a device offering it.
Two participants disliked \methodname when it switched to the \textit{static-large} UI because it reduced the number of visible soundtracks and required more navigation. 
One participant felt less in control when \methodname was not personalised to their preferences (e.g., shrinking icons with increased distance and enlarging them when closer). Another highlighted the importance of context in accepting \methodname. P19 attributed their negative experience to being a first-time user, explaining that \emph{\say{maybe if I just got more used to it, it would be easier … for the first time, it felt a bit confusing}}.}

\subsubsection{\methodname Threshold Well-Received, but Movement Matters}
\revision{Nine participants explicitly appreciated the adaptation threshold; \emph{\say{it was not too fast, nor too slow, it was great}} (P21). However, two participants found it too sensitive when walking, as P15 noted, \emph{\say{when you are moving, it is too frequent in changing}}. One participant viewed the threshold positively in static postures, while another felt it may have contributed to their negative experience with \methodname.}

\subsubsection{How Adaptive UI Shapes User Behaviour?}
\revision{Eleven participants reported that \methodname did not alter their behaviour. They adjusted the phone during the interaction for healthy viewing distance ($n=1$), better visibility ($n=1$), or comfort ($n=3$). Three experienced minimal adaptation because they typically hold the device at a fixed distance. On the other hand, 13 participants reported behavioural changes, including intentionally adjusting the UI to find a suitable size ($n=2$), to enlarge the interface ($n=1$), to improve selection accuracy ($n=9$), or to do so out of curiosity and fun ($n=2$). 
One participant uniquely leveraged \methodname by holding the phone close to navigate the pages more quickly, then moving it away to select the soundtracks smoothly. Two participants also moved their phones during the experiment to improve eye-tracking accuracy.}

\subsubsection{Designing for Users: Preferred Ways to Adapt the UI} \label{sec:preferred_ways_to_adapt}
\revision{17 participants reported that the adaptation logic used in \methodname-based media player made sense, by scaling up the size of the UI elements as the distance increased (see Section \ref{sec:prototype}), as P22 noted, \emph{\say{I found that to be the most accurate way}}. 
Some participants suggested adapting the UI to mimic a zoom-in/zoom-out effect by enlarging UI elements as the distance decreased ($n=4$). Four recommended a setting to choose between static and \textit{adaptive} UI, or allowing the UI to switch modes depending on posture or context ($n=6$).
Some participants suggested manually disabling adaptation in certain contexts or postures ($n=2$), \emph{\say{As the phone knows that I am walking, or the phone is moving, I would rather have a setting to turn it [the adaptation] off.}} (P17). Some suggested locking the current UI size to prevent further changes ($n=1$), or personalising the UI size based on what felt most comfortable ($n=1$). 
Participants also proposed alternative adaptation strategies that did not rely on viewing distance ($n=4$). Suggestions included enlarging targets when focusing on them ($n=1$), adapting the UI based on the number of on-screen targets ($n=1$), adjusting the UI when users struggle to make a selection ($n=1$), and designing \textit{adaptive} UIs to better support users with impaired vision ($n=1$).}

\subsubsection{Situational Preferences for Static UI}
\revision{While four participants expressed an overall preference for the static UI, some specifically preferred it when walking ($n=3$) or when sitting and slouching ($n=1$). One participant explicitly favoured the static UI, noting that the \textit{static-medium}, matched to their typical viewing distance, was the most comfortable in terms of size and readability.}

\subsubsection{Who Could Benefit from UI Adaptation?}
\revision{Three participants highlighted the potential benefits of \methodname for different user groups: the general population ($n=1$), individuals with poor eyesight ($n=1$), people with disabilities ($n=1$), and older adults ($n=2$). As P10 noted, if a zoom-in/zoom-out adaptation were implemented, \emph{\say{I can definitely see how that could be useful for older people … who often hold their phone very close … so making the text bigger … is helpful for them.}} One participant also suggested that the \textit{adaptive} UI could discourage users from bringing the phone too close to their face, potentially reducing eye strain or negative impacts on eyesight.}

\section{Discussion \& Future work}

In this section, we discuss the results of both phases of the experiment and provide a discussion on evaluating the adaptive user interface based on the participants' experiences. 

\subsection{Effectiveness of the Gaze-Adaptive User Interface}
\revision{Findings from Phase 1 showed that} the \textit{adaptive} UI resulted in significantly faster task times compared to the \textit{static-large} UI, \revision{and faster navigation time than the \textit{static-medium} and \textit{static-small} UIs, at the farthest distance, i.e., 35-39 cm}. \revision{The number of errors when playing soundtracks significantly reduced with the \textit{adaptive} UI at the farthest distance, compared to the \textit{static-small}.}
When counting the frequency with which participants accidentally activated the play/pause control while navigating the soundtrack list during the hard task, both the \textit{adaptive} and \textit{static-medium} UIs caused significantly fewer errors than the \textit{static-small} UI.
\revision{Overall, the \textit{adaptive} benefited from the strengths of the interfaces it contains, leveraging the strength of its distance-responsive design. It was faster than the \textit{static-large}, which requires frequent navigations, and produced fewer errors than \textit{static-small}, which relies on smaller target sizes at all distances. The \textit{Adaptive} UI also outperformed the \textit{static-medium} and \textit{static-small} in navigation at the farthest distance, where it uses the large elements. Such improvements suggest that the \textit{adaptive} UI can provide advantages in both speed and accuracy over the static UI, particularly in aspects where it demonstrated superior performance.}

While the \textit{adaptive} and \textit{static-medium} UIs received positive feedback from participants when ranking their preference for \textit{Interface Type}, the \textit{static-small} was least preferred. It was also perceived to cause a significantly higher workload compared to the \textit{static-medium} and \textit{static-large} UIs.
During the interview after the ESM, the \textit{static-small} was also reported to negatively impact the user experience by seven participants when the \textit{adaptive} UI switched to it at a closer viewing distance, with one explicitly suggesting a minimum size to prevent UI elements from becoming too small.
The \textit{static-large} UI received positive feedback from several participants, who attributed their positive experience to the \textit{adaptive}UI switching to the \textit{static-large} UI at a further distance.
Given the positive feedback the \textit{static-medium} UI also received from participants, as was described by a participant as the most comfortable size, we suggest that adaptive UIs could incorporate elements of both the \textit{static-medium} and the \textit{static-large} UIs to achieve optimal performance. The adaptive interface could maintain a basic size of 4°, calculated at face-to-screen distances between 30 and 35 cm; i.e., \textit{static-medium}. It then adapts to increased distances above 35 cm without reducing its element sizes at decreased distances under 30 cm.
 
The differences in navigation time between the \textit{static-large} and \textit{adaptive} compared to both the \textit{static-medium} and \textit{static-small} elements were significant at the farthest distance, i.e., [35-39] cm. This could be due to the regions in which these navigation buttons were placed. Such findings highlight the need for \textit{static-large} elements as the face-to-screen distance increased \revision{, which the \textit{adaptive} UI incorporates at farthest distances used in the experiment, i.e., > 35 cm. This is particularly relevant for navigation buttons, which were placed at the bottom region of the screen, where prior work showed that tracking accuracy degrades in that region \cite{10.1145/3706598.3713092}}.

Our experiment revealed that a few participants did not enjoy the consequences of the \textit{static-large} UI, which required more navigation. 
To avoid such consequences when switching to \textit{static-large}, elements of the user interface could use targets of medium sizes to accommodate more items while partially adapting the UI where needed, such as in the bottom regions of the phone, where accuracy deteriorates \cite{10.1145/3706598.3713092, 10.1145/3025453.3025599}. 
Additionally, some participants highlighted the penalty of adaptation, which could lead to confusion when adjusting to the changes in the UI; for example, to find the soundtrack in the current adapted list. To mitigate such an impact, we suggest exploring adapting regions of the screen, where dynamic contents of the interfaces can remain static, such as the soundtrack in the list. In contrast, control buttons can be adapted to enhance performance, especially if such buttons are located at the top or bottom regions of the phone. \revision{These findings are based on our choice of a media player to examine the concept of \methodname. The media player consisted of menu items, paginated navigation and buttons, with a reasonable level of complexity for a gaze-based interface. Applying the concept to other UI layouts or applications, or to several other tasks with other UI elements, would extend our understanding of the usability of adapting UIs for dwell interfaces.}

\subsection{Why Not Just Make the Targets Bigger}
\revision{One could simply design large targets for dwell-based input; however, this comes at the cost of displaying fewer elements at once, as large elements take up more screen real estate \cite{10.1145/2414536.2414609}, pushing other elements off-screen. 
Our results showed that \textit{static-small} UI, which required less navigation, resulted in significantly faster task times than the \textit{static-large} UI. This may be a consequence of designing the UI elements to be large: the larger the elements, the fewer that can fit on the screen, leading to more off-screen items and increased navigation. 
These consequences led some participants to dislike the \textit{static-large} UI because of the increased navigation required, as P5 noted, they had to navigate more pages regardless of distance. Overall, this also led participants to rank it as their second least-preferred option. 
On the other hand, when the \textit{static-large} UI is incorporated into the \textit{adaptive} UI and used only at the furthest distance, it received positive feedback.
This is the benefit of the \textit{adaptive} approach, which adjusts target sizes based on viewing distance without compromising accuracy, enlarging them only when needed. 
At closer distances, elements remain relatively small, allowing more elements to be displayed without compromising accuracy, while at farther distances, elements are enlarged to preserve accuracy, though this necessarily reduces the number of on-screen elements and increases navigation.
However, the extra navigation required at farther distances did not increase overall task time and drew little criticism from participants. 
In our experiment, although participants had to navigate more pages with the \textit{adaptive} UI at the farthest distances, we found that \textit{adaptive} still resulted in significantly faster task times than \textit{static-large}, with no evidence for significant differences compared to \textit{static-small} and \textit{static-medium}.
Additionally, the \textit{adaptive} UI improved navigation time relative to \textit{static-small} and \textit{static-medium} at the farthest distance, suggesting that performance degrades as elements remain relatively small with increasing distance. With the adaptive approach, the target size is dynamically adjusted as needed, ensuring that tracking accuracy is preserved despite changes in viewing distance.}

\subsection{Impact of Walking on Preference for Adaptive Interfaces}
Our results indicated that posture impacted participants' perceptions of the adaptive UI. While walking, the logged data showed that the average number of UI switches was nearly double that of the following highest postures: sitting (chair) and slouching, averaging 22.5 switches compared to 11.3 switches. 
Such frequent changes resulted in 13 participants reporting less favouring the \textit{adaptive} UI while walking, when interviewed. 
Subjective experiences collected using ESM in various postures with \textit{adaptive} UI also showed that walking resulted in the lowest expectation in terms of interface resizing, the least control over changes in UI, and was perceived as less helpful and resulted in less satisfaction. Participants also rated the UI as more sensitive while walking compared to other postures besides standing. Some participants also suggest a customised setting to turn adaptation off while walking. However, very few participants liked the \textit{adaptive} UI while walking; this could be due to holding the phone at a further distance between them and the device, as mentioned by one participant, which resulted in using the \methodname-based media player between the \textit{static-medium} and \textit{static-large} UIs. Given that \textit{adaptive} UI was generally acceptable by participants, given the findings, we suggest that it can be accepted while walking if the threshold is adjusted to accommodate frequent body movement while walking. Users can adjust the threshold in settings or via a calibration process, and can be offered an option to turn the adaptation off when walking. \revision{While in phase 1 of the experiment, we systematically explored the performance of the \textit{adaptive} UI against static UIs in a controlled setup, our findings on postures are based solely on participants' experience and perception of \methodname from the second phase. This opens the venue for future research to systematically explore the feasibility and performance of \methodname across various contexts and postures, such as when walking or sitting \cite{10.1145/3025453.3025794, Chittaro2010}.}

\subsection{Toward Personalised Adaptation}
While participants generally accepted the \textit{adaptive} UI, they suggested incorporating ways to personalise its behaviour. This could be to preserve their sense of agency, as prior work showed \cite{10.1145/3706598.3713367}. 
In our experiment, one participant highlighted that they felt less in control when the UI was not personalised based on their preference. Personalisation includes automatically or manually switching adaptation on or off depending on postures or contexts. For example, users can be given the option to lock the current UI size or to switch to adaptive when in static posture. 
Current smartphones' built-in sensors, such as the accelerometer and gyroscope, can help detect user posture by analysing motion patterns and acceleration changes to distinguish between postures such as sitting and walking. To avoid overloading users with frequent UI changes that could cause frustration, we recommend continuous monitoring of the adaptive behaviour and offering users an action button to disable adaptation if desired.
Some participants also suggested other adaptation behaviour as explained in section \ref{sec:preferred_ways_to_adapt}. This includes enlarging the targets as the distance decreases, mimicking the zoom-in/zoom-out effect. Such logic emerged from the fact that some participants reported that they usually bring their phone closer to their face to see more clearly or to focus on specific content. Future work could explore such adaptation logic in terms of usability.

\subsection{Adaptation for Healthy Viewing Distance}
While prior work utilised user interface adaptation to mitigate the impact of situational impairments on touch selection and reading while walking \cite{10.1145/1409240.1409253, 10.1145/3706598.3713367}, in this work, we utilised adaptation to enhance the performance of gaze-based dwell selection during everyday interaction with a mobile device, as the distance between the users and device varies depending on their context and posture. While the \textit{adaptive} UI showed promising results, some participants noted the health benefits of holding the phone at a specific distance. One participant suggested that the \textit{adaptive} UI could mitigate the negative effects of holding the phone closer to the eyes by allowing users to see targets more clearly at a distance. Prior work reported that holding the phone at a close distance is harmful according to medical research \cite{10.1145/2785830.2785836}.
While they implemented a prototype that warns users when the distance between them and their phone falls below their Harmon distance, encouraging a healthy viewing distance, adapting the user interface based on distance presents a promising alternative. Instead of issuing warnings, the UI could dynamically adjust so that UI elements remain easily readable and selectable from an appropriate distance, subtly guiding users toward healthier interaction habits. Such UI behaviour could be promising in preventing digital eyestrain (DES) that is more common amongst non-presbyopes as they tend to use the phone at a shorter distance with smaller font sizes compared to presbyopes \cite{Naipal2025}.

\subsection{Guidelines for Gaze-based Adaptive User Interfaces on Handheld Mobile Devices}
Based on the findings, we recommend the following when designing an adaptive user interface for dwell input on mobile devices: 

\begin{enumerate}[label=\textbf{[\arabic*]}]
   \item We recommend designing the UI elements to maintain a target size of 4° calculated at a distance between [30-34] cm and then adapt the UI as the distance increases to maintain the same visual angle. We recommend avoiding adapting the UI to maintain the 4° when the viewing distance is less than 30 cm, as this will negatively impact the performance of dwell selection.
    \item \revision{Consider allowing} users to switch adaptation on and off when needed. Mobile settings can provide users with options to switch between static and adaptive UIs based on their detected activities or behaviours. 
    \item For users who prefer to use adaptation while walking, the adaptive threshold should be adjustable and customisable by users to accommodate frequent body movement.
    \item \revision{Consider using} our recommendation for adaptation as the default setting, and allow users to customise its behaviour to their preference, as this should improve their sense of agency.
    \item To avoid the consequences of adapting the entire UI at farther distances, where larger targets may push elements off-screen and increase navigation effort, consider adapting regions where accuracy typically deteriorates, such as the lower region of the phone. 
    \item Consider partial adaptation of the UI by excluding regions with dynamic content from adaptation if accuracy is not an issue in such regions. Adapting regions with dynamic contents may disrupt user orientation. For example, when a soundtrack list expands from three to four visible items, it may cause users to lose track of their intended target.
\end{enumerate}

\section{Conclusion}
We conducted a study in two phases to explore the usability of \methodname, a gaze-based adaptive user interface, using a prototype media player developed specifically for this purpose. \revision{\textbf{In Phase 1,}} we compared the performance of the adaptive UI against three static UIs that varied in size as baselines. \revision{We found that the adaptive UI benefited from its distance-responsive design, leveraging the advantages of the interfaces it contains. It was faster than the \textit{static-large} UI and produced fewer errors than the \textit{static-small} UI. It was ranked as the most preferred UI.}
\revision{\textbf{In Phase 2,}} by adapting the ESM approach, we measured users' perceptions of \methodname while using the media player prototype in six different postures. \revision{Participants mostly appreciated using the adaptive UI in static postures, while they mostly did not prefer it while walking. They} suggested some level of personalisation, such as a setting to switch adaptation on or off, or to personalise when adaptation is to be activated, depending on their context. Based on the findings, we discussed the results and proposed guidelines for designing context-aware adaptive UIs for dwell interfaces on handheld mobile devices.

\bibliographystyle{ACM-Reference-Format}
\bibliography{sample-base}

\end{document}